\documentclass[12pt]{article}

\usepackage{amsfonts,amssymb,verbatim,amsmath}
\usepackage{setspace}
\usepackage[margin=1.0in]{geometry}
\usepackage[longnamesfirst,authoryear]{natbib}
\usepackage{graphicx}
\usepackage{appendix}
\usepackage{hyperref}
\hypersetup{colorlinks=true,anchorcolor=blue,citecolor=black,linkcolor=[rgb]{0,0.0,0.75}} %

\usepackage{caption}
\newtheorem{proposition}{Proposition}

\newtheorem{remark}{Remark}

\def\QTR#1#2{{\csname#1\endcsname {#2}}}%
\def\limfunc#1{\mathop{\rm #1}}%

\usepackage[table]{xcolor} %
\definecolor{myEvenRowsColor}{rgb}{0.93,0.93,1}

\begin{document}

\title{\textsc{Marginal Effects for Probit and Tobit with Endogeneity}}
\author{Kirill S. \textsc{Evdokimov}\thanks{%
Universitat Pompeu Fabra and Barcelona School of Economics: \textsf{%
kirill.evdokimov@upf.edu}.} \and Ilze \textsc{Kalnina}\thanks{%
North Carolina State University: \textsf{ikalnin@ncsu.edu}.} \and Andrei 
\textsc{Zeleneev}\thanks{%
University College London: \textsf{a.zeleneev@ucl.ac.uk}.\tiny\newline $ $ }}
\date{This version: December 2024}
\maketitle

    \begin{abstract}

      When evaluating partial effects, it is important to distinguish between structural endogeneity and measurement errors. In contrast to linear models, these two sources of endogeneity affect partial effects differently in nonlinear models. We study this issue focusing on the Instrumental Variable (IV) Probit and Tobit models. We show that even when a valid IV is available, failing to differentiate between the two types of endogeneity can lead to either under- or over-estimation of the partial effects. We develop simple estimators of the bounds on the partial effects and provide easy to implement confidence intervals that correctly account for both types of endogeneity. We illustrate the methods in a Monte Carlo simulation and an empirical application.

      \bigskip

    \end{abstract}

\newpage

\section{Introduction}

Probit and Tobit are some of the most popular nonlinear models in applied
economics. When a covariate is endogenous, IV-Probit and IV-Tobit models can
be used for instrumental variable (IV) estimation of the coefficients (%
\citealp{SmithBlundell1986Ecta}, \citealp{RiversVuong1988JoE}).\footnote{%
For example, in Stata, these estimators are \textit{ivprobit} and \textit{%
ivtobit}.}

A covariate can be endogenous for two reasons. First, the covariate can be
correlated with the individual's unobserved characteristics (unobserved
heterogeneity). Second, mismeasurement of the covariate also results in
endogeneity (Errors-in-Variables, EiV). We will refer to these two types of
endogeneity as the structural\ endogeneity and the EiV. In many empirical
settings both sources of endogeneity need to be addressed simultaneously.

In empirical applications of nonlinear models, 
researchers are often interested in partial effects and related counterfactuals. The goal of this paper is to characterize the partial effects in the classic IV-Probit and IV-Tobit models, allowing for both types of endogeneity, and to emphasize the
importance of distinguishing between the two types. We provide the
expressions for the partial effects and average partial effects that
correctly account for the two kinds of endogeneity. Although the two sources
of endogeneity cannot be precisely distinguished using the observed data, we
use the constraints of the model to obtain bounds on the amounts of
endogeneity that can be attributed to each source. This allows us to
characterize sharp bounds on the true partial effects and average partial effects,
allowing for both types of endogeneity. We also provide simple estimators of
these bounds and corresponding valid confidence intervals that are easy to
calculate.

The primary objects of interest of this paper are the partial effects of the
covariates, rather than the regression coefficients (the coefficients on the
covariates). Denote by $X_{i}^{\ast }$ the potentially endogenous covariate, %
 and by $X_{i}$ its mismeasured observed version. Estimation of the coefficients on $%
X_{i}^{\ast }$ and other covariates is a simpler task than estimation of the
partial effects of $X_{i}^{\ast }$. In particular, in the IV-Probit,
IV-Tobit, and related models, to identify and estimate these coefficients it
is sufficient to simply consider $X_{i}$ as the endogenous regressor of
interest without needing to distinguish between the types of endogeneity.%
\footnote{%
In particular, \cite{SmithBlundell1986Ecta} and \cite{RiversVuong1988JoE}
simply consider $X_{i}$ as endogenous. Similarly, in a recent paper, \cite%
{ChesherKimRosen2023JoE} provide a sharp identified set for the coefficients
on the covariates in a Tobit model with endogeneity under weak assumptions.
The approaches of these papers implicitly allow for mismeasured covariates,
as long as the focus is only on the regression coefficients.} %

In nonlinear models, the need to differentiate between the two kinds of
endogeneity arises because structural endogeneity and EiV play different
roles. In particular, partial effects of covariates are averaged with
respect to the distribution of the individual unobserved heterogeneity. On
the other hand, one aims to remove the impact of the measurement errors,
since they are not properties of individuals but a deficiency in the
measurement process. The textbook treatment of the problem often focuses
only on the first type of endogeneity, treating endogeneity as purely
structural. When $X_{i}^{\ast }$ is mismeasured, the partial effect of $%
X_{i}^{\ast }$ in nonlinear models differs from the effect of $X_{i}$ one
would calculate using the standard formulas that assume the endogeneity is
purely structural.{}

To identify the partial effects of $X_{i}^{\ast }$ and other covariates one
needs to identify the distribution of the true unobserved heterogeneity not
contaminated by the measurement error. It turns out that this distribution
is only partially identified. Thus, even though the IV-Probit and IV-Tobit
methods consistently estimate the coefficients on all regressors regardless
of the sources of endogeneity, the effects of the covariates on the outcomes
are only partially identified. The width of the identified set depends on
how hard it is to disentangle structural endogeneity and EiV for the data at
hand. Importantly, we find that naively ignoring the distinction between the
two types of endogeneity can result in both under- and over-estimation of
the magnitude of the partial effects by these IV estimators.\footnote{\cite%
{Wooldridge2010Book}, page 586, alludes to the potential importance of the
sources of endogeneity for the partial effects in IV-Probit, but does not
elaborate.}

IV-Probit and IV-Tobit can be interpreted as control variable estimators.
Partial effects in general control variable models were considered by \cite%
{BlundellPowell2003Advances}, \cite{Chesher2003Ecta}, \cite%
{ImbensNewey2009Ecta}, and \cite{Wooldridge2005APE,Wooldridge2015JHR}, among
others. These control variable methods focus on structural endogeneity
exclusively but do not consider EiV. The problems of estimation with
structural endogeneity or measurement errors are studied by two large but
(mostly) distinct literatures in econometrics, see, e.g., \cite%
{Matzkin2013AnnRev} and \cite{Schennach2020HB-ME} for reviews. In nonlinear
models, accounting for both types of endogeneity is challenging, see, e.g., 
\cite{Schennach2022JEcLit}. Exceptions include \cite%
{AdusumilliOtsu2018ET,SongSchennachWhite2015QE,SchennachWhiteChalak2012JoE,HahnRidder2017JoE}%
. These papers obtain point identification results when the distribution of
the measurement error is either known or can be recovered from repeated
measurements using the lemma of \cite{Kotlarski1967}. Such datasets,
however, are relatively rare.

Control variable methods in nonlinear models typically require the endogenous variable to be continuously distributed (e.g., see \citealp{ImbensNewey2009Ecta}, for a discussion). This limitation also applies to our framework: the mismeasured endogenous variable $X_i$ is assumed to be continuously distributed. Other covariates can be discrete. Note that properly accounting for both types of endogeneity of $X_i$ is essential for characterizing ceteris paribus effects of all covariates, including the discrete ones.

The advantage of gaussian nonlinear models is their simplicity and
transparency, which makes them a convenient starting point in an empirical
analysis. Our approach in particular provides the researchers with a simple
way to gauge the importance of properly accounting for the two types of
endogeneity, which is essential given the ubiquity of both in economic
applications. In addition, for the settings where relaxing gaussianity is
important, we develop an extension of our approach that allows both the
first stage unobservables and the measurement errors to be non-gaussian.

The rest of the paper is organized as follows. The analysis of partial
effects in the Probit and Tobit models is virtually identical, thus we first
focus on Tobit in Sections \ref{sec:Model}-\ref{sec:Analysis}. Section~\ref%
{sec:Probit} then considers the Probit model. Section~\ref{sec:APEs} extends the analysis to cover
the average partial effects and other counterfactuals. Section~\ref{sec:MC} provides
some Monte Carlo simulation results. Section~\ref{sec:Empirical} presents an
empirical application. Section~\ref%
{sec:Relax Gauss} relaxes gaussianity assumptions.

\bigskip

\section{The Model}\label{sec:Model}

The Tobit model is often used for estimation of economic models with a
\textquotedblleft corner solution,\textquotedblright\ i.e., models where the
outcome variable $Y_{i}\,$\ is forced to be non-negative. The examples of
such dependent variables $Y_{i}$ include the amounts of charitable
contributions, hours worked, or monthly consumption of cigarettes.

First, consider the standard Tobit model with exogenous covariates and
without EiV:%
\begin{equation}
Y_{i}=m\left( \theta _{01}X_{i}^{\ast }+\theta _{02}^{\prime
}W_{i}+U_{i}^{\ast }\right) ,\text{\quad where }m\left( s\right) =\max (s,0),
\label{eq:dgp Y exog copy}
\end{equation}%
the individual unobserved heterogeneity $U_{i}^{\ast }$ has a normal
distribution $N\left( 0,\sigma _{U^{\ast }}^{2}\right) $ and is independent
from the covariates $X_{i}^{\ast }$ and $W_{i}$. We use the asterisk to
denote variables that will be affected by the EiV, as we explain in detail
below.{}

We collect the covariates in a vector $H_{i}^{\ast }=\left( X_{i}^{\ast
},W_{i}^{\prime }\right) ^{\prime }$, so (\ref{eq:dgp Y exog copy}) can be
written as 
\begin{equation*}
Y_{i}=m\left( \theta _{0}^{\prime }H_{i}^{\ast }+U_{i}^{\ast }\right)
,\qquad H_{i}^{\ast }=\left( X_{i}^{\ast },W_{i}^{\prime }\right) ^{\prime
},\quad \theta _{0}=\left( \theta _{01},\theta _{02}^{\prime }\right)
^{\prime }\text{.}
\end{equation*}
We denote the standard normal cumulative distribution and density functions
by $\Phi $ and $\phi $.\footnote{%
Most of the analysis in Sections~\ref{sec:Model} and \ref{sec:Analysis}
equally applies to the Probit model. For simplicity of exposition, we focus
on the Tobit model for the moment, and then discuss Probit in Section~\ref%
{sec:Probit}.}

In the Tobit model, one is usually interested in the partial effects
(marginal effects) of covariates $H_{i}^{\ast }$ on $\mathrm{E}\left(
Y_{i}|H_{i}^{\ast }\right) $ and $\mathrm{P}\left( Y_{i}>0|H_{i}^{\ast
}\right) $. For concreteness we consider partial effects of the continuously
distributed covariates.

The partial effect of the $j^{th}$ covariate on the mean {}$\mathrm{E}\left( Y_{i}|H_{i}^{\ast }=h\right) $ at a
given $h$ is %
\begin{equation}
PE_{j}^{\text{Tob}}\left( h\right) =\frac{\partial }{\partial h_{j}}\int
m\left( \theta _{0}^{\prime }h+u\right) f_{U^{\ast }}\left( u\right) du=\Phi
\left( \frac{\theta _{0}^{\prime }h}{\sigma _{U^{\ast }}}\right) \theta
_{0j}.  \label{eq:PE_Tobit}
\end{equation}%
The partial effect of the $j^{th}$ covariate on the probability $P\left(
Y_{i}>0|H_{i}^{\ast }=h\right) $ is %
\begin{equation}
PE_{j}^{\Pr }\left( h\right) =\frac{\partial }{\partial h_{j}}\int 1\left\{
\theta _{0}^{\prime }h+u>0\right\} f_{U^{\ast }}\left( u\right) du=\phi
\left( \frac{\theta _{0}^{\prime }h}{\sigma _{U^{\ast }}}\right) \frac{%
\theta _{0j}}{\sigma _{U^{\ast }}}.  \label{eq:PE_Probit}
\end{equation}%
These formulas for the $PE_{j}$ are standard, see, e.g., \cite%
{Wooldridge2010Book}, for detailed calculations. Most often one considers
the partial effects at the means of the covariates $h=E\left[ H_{i}^{\ast }%
\right] $ or partial effects averaged with respect to the distribution of $H_i^*$.

When $X_{i}^{\ast }$ is correlated with $U_{i}^{\ast }$ and we observe data $%
\left( Y_{i},X_{i}^{\ast },W_{i},Z_{i}\right) $, the IV-Tobit model can be
estimated using instrumental variables $Z_{i}$, as proposed by \cite%
{SmithBlundell1986Ecta}, \cite{Newey1987}, and \cite{RiversVuong1988JoE}.
Assume that%
\begin{eqnarray}
Y_{i} &=&m\left( \theta _{01}X_{i}^{\ast }+\theta _{02}^{\prime
}W_{i}+U_{i}^{\ast }\right) ,\text{\quad }m\left( s\right) =\max (s,0),
\label{eq:dgp Y true} \\
X_{i}^{\ast } &=&\pi _{01}^{\prime }Z_{i}+\pi _{02}^{\prime
}W_{i}+V_{i}^{\ast },\quad \pi _{01}^{\prime }\neq 0,
\label{eq:dgp Xstar gen}
\end{eqnarray}%
where $V_{i}^{\ast }$ is a normal random variable, possibly correlated with $%
U_{i}^{\ast }$,%
\begin{equation}
\left( 
\begin{array}{c}
U_{i}^{\ast } \\ 
V_{i}^{\ast }%
\end{array}%
\right) \sim N\left( \left( 
\begin{array}{c}
0 \\ 
0%
\end{array}%
\right) ,%
\begin{pmatrix}
\sigma _{U^{\ast }}^{2} & \sigma _{U^{\ast }V^{\ast }} \\ 
\sigma _{U^{\ast }V^{\ast }} & \sigma _{V^{\ast }}^{2}%
\end{pmatrix}%
\right) ,  \label{eq:dgp UV gen}
\end{equation}%
and $\left( U_{i}^{\ast },V_{i}^{\ast }\right) $ is independent from $\left(
Z_{i},W_{i}\right) $. In this model $X_{i}^{\ast }$ is continuously
distributed.

The IV-Tobit model in (\ref{eq:dgp Y true})-(\ref{eq:dgp UV gen}) can be
estimated using a random sample of $\left( Y_{i},X_{i}^{\ast
},W_{i},Z_{i}\right) $ in two steps, see, e.g., \cite{Wooldridge2010Book}.
First, one estimates $V_{i}^{\ast }$ in equation (\ref{eq:dgp Xstar gen}) by
the residuals $\hat{V}_{i}^{\ast }$ in the regression of $X_{i}^{\ast }$ on $%
\left( W_{i},Z_{i}\right) $. Note that we can write $U_{i}^{\ast
}=e_{i}^{\ast }+\theta _{V^{\ast }}V_{i}^{\ast }$, where $\theta _{V^{\ast
}}\equiv \sigma _{U^{\ast }V^{\ast }}/\sigma _{V^{\ast }}^{2}$, and $%
e_{i}^{\ast }$ is independent of $Z_{i}$, $W_{i}$, and $V_{i}^{\ast }$ (and
hence of $X_{i}^{\ast }$). Then, one estimates the standard Tobit model 
\begin{equation*}
Y_{i}=m\left( \theta _{01}X_{i}^{\ast }+\theta _{02}^{\prime }W_{i}+\theta
_{V^{\ast }}V_{i}^{\ast }+e_{i}^{\ast }\right) ,
\end{equation*}%
where $V_{i}^{\ast }$ are replaced by their estimates $\hat{V}_{i}^{\ast }$.
(Alternatively, the two steps can be combined and all of the parameters can
be estimated simultaneously by the Maximum Likelihood Estimator.) The reason
this approach works is that equation (\ref{eq:dgp Xstar gen}) creates a
control variable $V_{i}^{\ast }$, and the inclusion of $V_{i}^{\ast }$ in
the above equation makes $X_{i}^{\ast }$ exogenous. {}

To estimate the partial effects, one would plug the estimates $\widehat{%
\theta }$ and $\widehat{\sigma }_{U^{\ast }}^{2}$ into equations~(\ref%
{eq:PE_Tobit})-(\ref{eq:PE_Probit}) in place of $\theta _{0}$ and $\sigma
_{U^{\ast }}^{2}$.

So far we were assuming that the data has no measurement errors. We now
allow $X_{i}^{\ast }$ to be mismeasured, i.e., that instead of $X_{i}^{\ast
} $ we observe its noisy measurement $X_{i}$:%
\begin{equation}
X_{i}=X_{i}^{\ast }+\varepsilon _{i},\qquad \varepsilon _{i}\sim N\left(
0,\sigma _{\varepsilon }^{2}\right) .  \label{eq:ME}
\end{equation}%
We assume that $\varepsilon _{i}\perp \left( U_{i}^{\ast },V_{i}^{\ast
},W_{i},Z_{i}\right) $, i.e., the measurement error is classical. %
 The normality assumption simplifies the analysis but it is not crucial. We relax it in Section~\ref{sec:Relax Gauss}.

Note that the researcher's object of interest has not changed: the goal is
to estimate the partial effects defined in equations~(\ref{eq:PE_Tobit})-(%
\ref{eq:PE_Probit}). The structural endogeneity and measurement errors are
difficulties that an estimation procedure needs to overcome. In particular,
note that we are interested in estimation of the effect of $X_{i}^{\ast }$
and \emph{not} in the effect of the error-laden $X_{i}$.\footnote{%
This is similar to the linear regression settings, where one would be
interested in the effect of $X_{i}^{\ast }$ on $Y_{i}$. The slope
coefficient in the OLS\ regression of $Y_{i}$ on $X_{i}$ is not the object
of interest because it is subject to the attenuation bias due to the EiV
(and also possibly due to the endogeneity of $X_{i}^{\ast }$).}

\bigskip

\section{Analysis of the Model}\label{sec:Analysis}

First, we use the model in equations (\ref{eq:dgp Y true})-(\ref{eq:ME}) to
obtain the model in terms of the observable $X_{i}$. Since $X_{i}^{\ast
}=X_{i}-\varepsilon _{i}$, we can rewrite (\ref{eq:dgp Y true}) as 
\begin{eqnarray*}
Y_{i} &=&m\left( \theta _{01}X_{i}^{\ast }+\theta _{02}^{\prime
}W_{i}+U_{i}^{\ast }\right) =m\left( \theta _{01}X_{i}+\theta _{02}^{\prime
}W_{i}-\theta _{01}\varepsilon _{i}+U_{i}^{\ast }\right) \\
&=&m\left( \theta _{01}X_{i}+\theta _{02}^{\prime }W_{i}+U_{i}\right) ,
\end{eqnarray*}%
where $U_{i}\equiv U_{i}^{\ast }-\theta _{01}\varepsilon _{i}$. Let $%
V_{i}\equiv V_{i}^{\ast }+\varepsilon _{i}$. The model in equations (\ref%
{eq:dgp Y true})-(\ref{eq:ME}) can be written as%
\begin{eqnarray}
Y_{i} &=&m\left( \theta _{01}X_{i}+\theta _{02}^{\prime }W_{i}+U_{i}\right) ,
\label{eq:dgp Y wo stars gen} \\
X_{i} &=&\pi _{01}^{\prime }Z_{i}+\pi _{02}^{\prime }W_{i}+V_{i},
\label{eq:dgp X wo stars gen} \\
\left( 
\begin{array}{c}
U_{i} \\ 
V_{i}%
\end{array}%
\right) &\sim &N\left( \left( 
\begin{array}{c}
0 \\ 
0%
\end{array}%
\right) ,%
\begin{pmatrix}
\sigma _{U}^{2} & \sigma _{UV} \\ 
\sigma _{UV} & \sigma _{V}^{2}%
\end{pmatrix}%
\right) .  \label{eq:dgp UV wo stars gen}
\end{eqnarray}%
The definitions of $U_{i}$ and $V_{i}$ imply that%
\begin{equation}
\sigma _{U}^{2}=\sigma _{U^{\ast }}^{2}+\theta _{01}^{2}\sigma _{\varepsilon
}^{2},\quad \sigma _{V}^{2}=\sigma _{V^{\ast }}^{2}+\sigma _{\varepsilon
}^{2},\quad \sigma _{UV}=\sigma _{U^{\ast }V^{\ast }}-\theta _{01}\sigma
_{\varepsilon }^{2}.  \label{eq:sigma2 w and wo stars}
\end{equation}

Note that variables $X_{i},U_{i},V_{i}$ are the analogs of the true
variables $X_{i}^{\ast },U_{i}^{\ast },V_{i}^{\ast }$ that arise due to the
measurement errors $\varepsilon _{i}$. In the absence of measurement errors,
i.e., when $\varepsilon _{i}=0$, we have $X_{i}=X_{i}^{\ast }$, $%
U_{i}=U_{i}^{\ast }$, $V_{i}=V_{i}^{\ast }.$

The model in equations~(\ref{eq:dgp Y wo stars gen})-(\ref{eq:dgp UV wo
stars gen}) can be estimated by MLE or using the control variable two-step
approach described earlier. Specifically, both approaches will consistently
estimate parameters $\theta _{0}$ and the covariance matrix of the
unobservables in equation (\ref{eq:dgp UV wo stars gen}), i.e., $\sigma
_{U}^{2}$, $\sigma _{UV}$, and $\sigma _{V}^{2}$. 

Note that because the
model is nonlinear, the marginal effects defined in equations~(\ref%
{eq:PE_Tobit})-(\ref{eq:PE_Probit}) depend not only on $\theta _{0}$ but
also on $\sigma _{U^{\ast }}^{2}$. Thus, even though the available data $%
\left(Y_{i},X_{i},W_{i},Z_{i}\right) $ allows immediately estimating $\theta
_{0}$, we cannot obtain the marginal effects because we do not know $%
\sigma_{U^{\ast }}^{2}$. Naively using an estimate of $\sigma _{U}^{2}$ in
place of $\sigma _{U^{\ast }}^{2}$ would lead to a biased estimate of the
partial effects, since $\sigma _{U}^{2}\geq \sigma _{U^{\ast }}^{2}$, as
implied by equation~(\ref{eq:sigma2 w and wo stars}).

The problem with identifying $\sigma _{U^{\ast }}^{2}$ is that the data only
allows identification of the $3$ parameters $\sigma _{U}^{2}$, $\sigma _{UV}$%
, and $\sigma _{V}^{2}$. However, the distribution of the true $\left(
U_{i}^{\ast },V_{i}^{\ast },\varepsilon _{i}\right) $ is governed by $4$
parameters: $\sigma _{U^{\ast }}^{2}$, $\sigma _{U^{\ast }V^{\ast }}$, $%
\sigma _{V^{\ast }}^{2}$, and $\sigma _{\varepsilon }^{2}$. Thus, one cannot
uniquely determine these $4$ parameters from the $3$ equations~(\ref%
{eq:sigma2 w and wo stars}). In other words, models with different values of 
$\sigma _{\varepsilon }^{2}$ are observationally equivalent: they correspond
to identical distributions of the observables $\left(
Y_{i},X_{i},W_{i},Z_{i}\right) $ even though they imply different values of
true $\sigma _{U^{\ast }}^{2}$. Thus, one cannot uniquely determine (i.e.,
point-identify) $\sigma _{U^{\ast }}^{2}$ from the data $\left(
Y_{i},X_{i},W_{i},Z_{i}\right) $. Correspondingly, one cannot point-identify
the partial effects, which depend on $\sigma _{U^{\ast }}^{2}$.

Equations~(\ref{eq:sigma2 w and wo stars}) provide restrictions on $\sigma
_{U^{\ast }}^{2}$, which we will use to provide bounds on the possible
values of true $\sigma _{U^{\ast }}^{2}$, and hence on the values of the
partial effects.

\paragraph{Bounds on $\protect\sigma _{U^{\ast }}^{2}$ \label%
{sec:sig_Ustar_bounds}}

From equations~(\ref{eq:sigma2 w and wo stars}) the upper bound on $\sigma
_{U^{\ast }}^{2}$ is $\sigma _{U^{\ast }}^{2}\leq \sigma _{U}^{2}$. We now
obtain the lower bound on $\sigma _{U^{\ast }}^{2}$. In particular, we look
to find the smallest $\sigma _{U^{\ast }}^{2}$ that satisfies equations~(\ref%
{eq:sigma2 w and wo stars}), Cauchy-Schwarz inequality $\sigma _{U^{\ast
}V^{\ast }}^{2}\leq \sigma _{U^{\ast }}^{2}\sigma _{V^{\ast }}^{2}$, and the
non-negativity constraints $\sigma _{U^{\ast }}^{2}\geq 0$, $\sigma
_{V^{\ast }}^{2}\geq 0$, and $\sigma _{\varepsilon }^{2}\geq 0$. Let $\rho
_{UV}=\mathrm{corr}\left( U_{i},V_{i}\right) $.

\begin{proposition}
\label{prop:Ustar bounds}Suppose $\left\vert \rho _{UV}\right\vert <1$ in
model~(\ref{eq:dgp Y wo stars gen})-(\ref{eq:dgp UV wo stars gen}). Then the
sharp identified set for $\sigma_{U^*}^2$ is given by 
\begin{equation*}
\sigma _{U^{\ast }}^{2}\in \left[ \underline{\sigma }_{U^{\ast }}^{2},\sigma
_{U}^{2}\right] ,
\end{equation*}%
where%
\begin{equation}
\underline{\sigma }_{U^{\ast }}^{2}\equiv \max \left\{ \frac{\left( \theta
_{01}\sigma _{UV}+\sigma _{U}^{2}\right) ^{2}}{\sigma _{V}^{2}\theta
_{01}^{2}+2\sigma _{UV}\theta _{01}+\sigma _{U}^{2}},\ \sigma
_{U}^{2}-\theta _{01}^{2}\sigma _{V}^{2}\right\} .  \label{eq:bound s2Us alt}
\end{equation}
\end{proposition}

Proposition \ref{prop:Ustar bounds} provides the bounds in terms of the
quantities that can be estimated using the data $\left(
Y_{i},X_{i},W_{i},Z_{i}\right) $. Condition $\left\vert \rho
_{UV}\right\vert <1$ guarantees that the denominator in the fraction above
is positive. The proof of Proposition \ref{prop:Ustar bounds} also provides
bounds on $\sigma _{U^{\ast }V^{\ast }}$ and $\sigma _{\varepsilon }^{2}$.

\paragraph{Correct Partial Effects\label{sec:correct PE}}

We now use the bounds on $\sigma _{U^{\ast }}^{2}$ from Proposition~%
\ref{prop:Ustar bounds} to obtain the bounds on the partial effects, in
terms of the parameters that can be recovered from data.{} For simplicity and concreteness of exposition, we first consider partial effects evaluated at some fixed values of covariates. We consider average partial effects and other counterfactuals of interest in Section~\ref{sec:APEs}.

{}For a given $\sigma _{U^{\ast }}^{2}$,
the partial effects for the $j^{th}$ covariate are defined as in equations (%
\ref{eq:PE_Tobit})-(\ref{eq:PE_Probit}),%
\begin{equation}
PE_{j}^{\text{Tob}}\left( h,\sigma _{U^{\ast }}^{2}\right) =\Phi \left( 
\frac{\theta _{0}^{\prime }h}{\sigma _{U^{\ast }}}\right) \theta _{0j}\text{
\ \ and \ \ }PE_{j}^{\Pr }\left( h,\sigma _{U^{\ast }}^{2}\right) =\phi
\left( \frac{\theta _{0}^{\prime }h}{\sigma _{U^{\ast }}}\right) \frac{%
\theta _{0j}}{\sigma _{U^{\ast }}}.  \label{eq:PE_ProbitTobit_sigStar}
\end{equation}

The lower and upper bounds for partial effects for the $j^{th}$ covariate, $%
PE_{j}\left( h\right) $, are computed as 
\begin{equation}
\min_{\sigma _{U^{\ast }}^{2}\in \left[ \underline{\sigma }_{U^{\ast
}}^{2},\sigma _{U}^{2}\right] }PE_{j}\left( h,\sigma _{U^{\ast }}^{2}\right) 
\text{\quad and\quad }\max_{\sigma _{U^{\ast }}^{2}\in \left[ \underline{%
\sigma }_{U^{\ast }}^{2},\sigma _{U}^{2}\right] }PE_{j}\left( h,\sigma
_{U^{\ast }}^{2}\right) .  \label{eq:bounds_PE_generic}
\end{equation}

Function $PE_{j}^{\text{Tob}}\left( h,\sigma _{U^{\ast }}^{2}\right) $ in (%
\ref{eq:PE_ProbitTobit_sigStar}) is a monotone function of $\sigma _{U^{\ast
}}^{2}$, so the minimum and maximum in equation (\ref{eq:bounds_PE_generic})
are achieved on the boundaries of interval $\left[ \underline{\sigma }%
_{U^{\ast }}^{2},\sigma _{U}^{2}\right] $.

Function $PE_{j}^{\Pr }\left( h,\sigma _{U^{\ast }}^{2}\right) $ in
equation~(\ref{eq:PE_ProbitTobit_sigStar}) is not monotone in $\sigma
_{U^{\ast }}^{2}$, but the bounds in equation (\ref{eq:bounds_PE_generic})
for $PE_{j}^{\Pr }\left( h\right) $ can also be simplified. The minimum and
maximum over $\sigma _{U^{\ast }}^{2}\in \left[ \underline{\sigma }_{U^{\ast
}}^{2},\sigma _{U}^{2}\right] $ can be attained only at $\sigma _{U^{\ast
}}^{2}=\underline{\sigma }_{U^{\ast }}^{2}$, at $\sigma _{U^{\ast
}}^{2}=\sigma _{U}^{2}$, and, when $\left( \theta _{0}^{\prime }h\right)
^{2}\in \left[ \underline{\sigma }_{U^{\ast }}^{2},\sigma _{U}^{2}\right] $,
at $\sigma _{U^{\ast }}^{2}=\left( \theta _{0}^{\prime }h\right) ^{2}$.
Thus, one only needs to evaluate $PE_{j}^{\Pr }\left( h,\sigma _{U^{\ast
}}^{2}\right) $ at these $2$ or $3$ points to calculate the minimum and
maximum in equation~(\ref{eq:bounds_PE_generic}).

Since $\sigma _{U}\geq \sigma _{U^{\ast }}$, naively using $\sigma _{U}$
instead of $\sigma _{U^{\ast }}$ when calculating $PE_{j}^{\text{Tob}}\left(
h\right) $, would lead to attenuation bias when $\theta _{0}^{\prime }h>0$,
but would bias $PE_{j}^{\text{Tob}}\left( h\right) $ away from zero when $%
\theta _{0}^{\prime }h<0$, i.e., the EiV would make naive $PE_{j}^{\text{Tob}%
}\left( h,\sigma _{U}^{2}\right) $ over-estimate the partial effects $%
PE_{j}^{\text{Tob}}\left( h\right) $ in the latter case. Likewise, for the
probability, naively using $PE_{j}^{\Pr }\left( h,\sigma _{U}^{2}\right) $
can both under- and over-estimate the true partial effect $PE_{j}^{\Pr
}\left( h\right) $.

\paragraph{Estimation}

Using the standard two-step or MLE approaches described in Section \ref%
{sec:Model}, one obtains the estimates of $\theta _{0}$, $\sigma _{U}^{2}$, $%
\sigma _{V}^{2}$, and $\sigma _{UV}$ (and of their variance-covariance
matrix for inference). Then, from equation (\ref{eq:bound s2Us alt}) one
obtains the estimate of $\underline{\sigma }_{U^{\ast }}^{2}$.

For a given value of $\sigma _{U^{\ast }}^{2}$, the estimated partial
effects would be%
\begin{equation}
\widehat{PE}_{j}^{\text{Tob}}\left( h,\sigma _{U^{\ast }}^{2}\right) =\Phi
\left( \frac{\widehat{\theta }^{\prime }h}{\sigma _{U^{\ast }}}\right) 
\widehat{\theta }_{j}\text{ \ \ and \ \ }\widehat{PE}_{j}^{\Pr }\left(
h,\sigma _{U^{\ast }}^{2}\right) =\phi \left( \frac{\widehat{\theta }%
^{\prime }h}{\sigma _{U^{\ast }}}\right) \frac{\widehat{\theta }_{j}}{\sigma
_{U^{\ast }}}.  \label{eq:hat PE-ProbitTobit(sigUs2)}
\end{equation}%
 Then, the estimated bounds on $%
PE_{j}\left( h\right) $ are %
\begin{equation}
\min_{v\in \left[ \widehat{\underline{\sigma }}_{U^{\ast }}^{2},\widehat{%
\sigma }_{U}^{2}\right] }\widehat{PE}_{j}\left( h,v\right) \text{\quad
and\quad }\max_{v\in \left[ \widehat{\underline{\sigma }}_{U^{\ast }}^{2},%
\widehat{\sigma }_{U}^{2}\right] }\widehat{PE}_{j}\left( h,v\right) ,
\label{eq:bounds_hat_PE_generic}
\end{equation}%
where the minimum and maximum are easily computed using univariate numerical
optimization. For the partial effects in equation (\ref{eq:hat
PE-ProbitTobit(sigUs2)}), these extrema can also be computed as described
under equation (\ref{eq:bounds_PE_generic}).

For example, one often considers the partial effects at the mean values of
covariates taking $h=(\overline{X},\overline{W}^{\prime })^{\prime }$, where 
$\overline{X}$ and $\overline{W}$ are the sample averages.  Note
that $\mathrm{E}\left[ X_{i}\right] =\mathrm{E}\left[ X_{i}^{\ast }\right] $.

\paragraph{Inference}

To provide a simple method for inference about the partial effects, we adopt
a Bonferroni approach (e.g., \citealp{McCloskey2017JoE}). This approach
allows us to avoid computational challenges that often arise in the context
of subvector inference in partially identified models. The construction of a 
$1-\alpha $ confidence interval for a partial effect $PE_{j}(h)$ proceeds in
two steps:

\begin{enumerate}
\item Pick $\alpha _{1}\in (0,\alpha )$ and construct $CI_{1-\alpha
_{1}}^{\sigma _{U^{\ast }}^{2}}$, a $1-\alpha _{1}$ confidence interval for $%
\sigma _{U^{\ast }}^{2}$, based on the bounds provided in Proposition~%
\ref{prop:Ustar bounds}.

\item Construct a $1-\alpha $ confidence interval for $PE_{j}(h)$ as the
union $CI_{1-\alpha }^{PE_{j}(h)}=\bigcup_{\sigma _{U^{\ast }}^{2}\in
CI_{1-\alpha _{1}}^{\sigma _{U^{\ast }}^{2}}}CI_{1-\left( \alpha -\alpha
_{1}\right) }^{PE_{j}(h)}\left( \sigma _{U^{\ast }}^{2}\right) $, where $%
CI_{1-\left( \alpha -\alpha _{1}\right) }^{PE_{j}(h)}\left( \sigma _{U^{\ast
}}^{2}\right) $ is a standard $1-(\alpha -\alpha _{1})$ confidence interval
for $PE_{j}(h)$ based on $\widehat{PE}_{{j}}\left( h,\sigma _{U^{\ast
}}^{2}\right) $ in equation~(\ref{eq:hat PE-ProbitTobit(sigUs2)}) for a
given $\sigma _{U^{\ast }}^{2}$.
\end{enumerate}

We now provide the implementation details for each step.

\smallskip

\noindent \textit{Step 1}. The confidence interval for $\sigma _{U^{\ast
}}^{2}$ is constructed based on the bounds given in Proposition~\ref%
{prop:Ustar bounds}. As the upper bound, we take $\hat{\sigma}%
_{U}^{2}+z_{1-\alpha _{1}/2}\times s_{\hat{\sigma}_{U}^{2}}$, where $s_{\hat{%
\sigma}_{U}^{2}}$ is the standard error of $\hat{\sigma}_{U}^{2}$, and $%
z_{1-\alpha _{1}/2}$ is the $1-\alpha _{1}/2$ quantile of the standard
normal distribution. The lower bound is based on $\underline{\hat{\sigma}}%
_{U^{\ast }}^{2}=\max \{\hat{\xi}_{1},\hat{\xi}_{2}\}$, where $\hat{\xi}_{1}$
and $\hat{\xi}_{2}$ are the plug-in estimators of the two terms on the right
hand side of equation~(\ref{eq:bound s2Us alt}). Note that $\hat{\xi}_{1}$
and $\hat{\xi}_{2}$ are (generally) jointly asymptotically normal and their
asymptotic variance-covariance matrix can be computed using the delta
method. Then, as the lower bound of $CI_{1-\alpha _{1}}^{\sigma _{U^{\ast
}}^{2}}$, we take $\max \{\hat{\xi}_{1}-c_{1-\alpha _{1}/2}\times s_{\hat{\xi%
}_{1}},\hat{\xi}_{2}-c_{1-\alpha _{1}/2}\times s_{\hat{\xi}_{2}}\}$. Here $%
s_{\hat{\xi}_{1}}$ and $s_{\hat{\xi}_{2}}$ are the standard errors of $\hat{%
\xi}_{1}$ and $\hat{\xi}_{2}$, and $c_{1-\alpha _{1}/2}$ is the $1-\alpha
_{1}/2$ quantile of $\max \{\eta _{1},\eta _{2}\}$, where $\left( \eta
_{1},\eta _{2}\right) $ are jointly normal with unit variances and
correlation $\hat{\rho}_{\hat{\xi}_{1},\hat{\xi}_{2}}$, and $\hat{\rho}_{%
\hat{\xi}_{1},\hat{\xi}_{2}}$ is an estimator of the correlation between $%
\hat{\xi}_{1}$ and $\hat{\xi}_{2}$ (e.g., see \citealp{RomanoWolf2005Ecta}).
By a standard argument, the confidence interval for $\sigma _{U^{\ast }}^{2}$
given by 
\begin{equation*}
CI_{1-\alpha _{1}}^{\sigma _{U^{\ast }}^{2}}=\left[ \max \left\{ \hat{\xi}%
_{1}-c_{1-\alpha _{1}/2}\times s_{\hat{\xi}_{1}},\hat{\xi}_{2}-c_{1-\alpha
_{1}/2}\times s_{\hat{\xi}_{2}}\right\} ,\hat{\sigma}_{U}^{2}+z_{1-\alpha
_{1}/2}\times s_{\hat{\sigma}_{U}^{2}}\right]
\end{equation*}%
has asymptotic coverage at least $1-\alpha _{1}$ for the true $\sigma
_{U^{\ast }}^{2}$. In the numerical illustrations we take $\alpha
_{1}=\alpha /10$.

\smallskip

\noindent \textit{Step 2}. First, the standard $CI_{1-\left( \alpha -\alpha
_{1}\right) }^{PE_{j}(h)}\left( \sigma _{U^{\ast }}^{2}\right) $ is $\left[
l_{1-\left( \alpha -\alpha _{1}\right) }^{PE_{j}(h)}\left( \sigma _{U^{\ast
}}^{2}\right) ,u_{1-\left( \alpha -\alpha _{1}\right) }^{PE_{j}(h)}\left(
\sigma _{U^{\ast }}^{2}\right) \right]$ constructed by adding and
subtracting $z_{1-(\alpha -\alpha _{1})/2}\times s_{\widehat{PE}%
_{j}(h,\sigma _{U^{\ast }}^{2})}$ from $\widehat{PE}_{j}(h,\sigma _{U^{\ast
}}^{2})$. The standard error $s_{\widehat{PE}_{j}(h,\sigma _{U^{\ast
}}^{2})} $ of $\widehat{PE}_{j}(h,\sigma _{U^{\ast }}^{2})$ can be computed
using the delta method. Then we can construct $CI_{1-\alpha }^{PE_{j}(h)}$
as 
\begin{equation*}
CI_{1-\alpha }^{PE_{j}(h)} = \left[\min_{\sigma _{U^{\ast }}^{2}\in
CI_{1-\alpha _{1}}^{\sigma _{U^{\ast }}^{2}}}l_{1-\left( \alpha -\alpha
_{1}\right) }^{PE_{j}(h)}\left( \sigma _{U^{\ast }}^{2}\right) ,
\max_{\sigma _{U^{\ast }}^{2}\in CI_{1-\alpha _{1}}^{\sigma _{U^{\ast
}}^{2}}}u_{1-\left( \alpha -\alpha _{1}\right) }^{PE_{j}(h)}\left( \sigma
_{U^{\ast }}^{2}\right)\right] ,
\end{equation*}%
where the minimum and maximum are easily calculated using univariate
numerical optimization over $\sigma _{U^{\ast }}^{2}$. By the standard
Bonferroni argument, the confidence interval $CI_{1-\alpha }^{PE_{j}(h)}$
has asymptotic coverage of at least $1-\alpha $ for the true partial effect $%
PE_{j}(h)$.

\smallskip

The constructed confidence interval is asymptotically valid as long as (i)
the first step confidence interval $CI_{1-\alpha _{1}}^{\sigma _{U^{\ast
}}^{2}}$ covers the true $\sigma _{U^{\ast }}^{2}$ with probability at least 
$1-\alpha _{1}$ asymptotically, and (ii) the delta method applies to $%
\widehat{PE}_{j}(h,\sigma _{U^{\ast }}^{2})$ for the true $\sigma _{U^{\ast
}}^{2}$. Both conditions are satisfied provided that the true $\sigma
_{U^{\ast }}^{2}$ is bounded away from zero. Note that in this case $%
CI_{1-\alpha _{1}}^{\sigma _{U^{\ast }}^{2}}$ is valid even if $\theta
_{01}\sigma _{UV}+\sigma _{U}^{2}$ is equal to (or local to) zero, which
implies that $CI_{1-\alpha }^{PE_{j}(h)}$ is also valid.

\begin{remark}[On the classical measurement error assumption]
  Our method relies on the control variable approach to identifying $\theta_0$ (and the other reduced form parameters), which requires $(U_i,V_i) \perp Z_i$. Full independence is a standard requirement needed for point identification in nonlinear models for control variable methods (e.g., \citealp{ImbensNewey2009Ecta}). Since $U_i = U_i^* - \theta_{01} \varepsilon_i$ and $V_i = V_i^* + \varepsilon_i$, the control variable approach requires $\varepsilon_i$ to be independent of $Z_i$. Full independence is necessary even for the standard IV-Tobit and IV-Probit models of \citet{SmithBlundell1986Ecta} and \citet{RiversVuong1988JoE}: when $X_i^*$ is mismeasured, their assumptions implicitly require $\varepsilon_i \perp (Z_i, W_i)$. On the other hand, as shown in Section~\ref{sec:Relax Gauss}, the normality assumption $\varepsilon_i \sim N (0,\sigma_\varepsilon^2)$ is not crucial for identification and can be relaxed.
\end{remark}

\section{Probit}\label{sec:Probit}

IV-Probit model is the same as IV-Tobit except $m\left( s\right) =1\left\{ s>0\right\} $ in equation (\ref{eq:dgp Y true}). Since Probit is a binary outcome model, in equation~(\ref{eq:dgp Y wo stars gen}) one needs to impose a scale normalization, for example, $\left\Vert \theta _{0}\right\Vert =1$ or $\sigma_{U}^2 = 1$. For both of those normalizations, $\theta_0$ and the other reduced form parameters are point identified, and the sharp bounds for $\sigma_{U^*}^2$ are given by Proposition~\ref{prop:Ustar bounds} as before.

For Probit, we are interested in the partial effects of
covariates on the probability of $Y_{i}=1$, which are given by $PE_{j}^{\Pr
}\left( h\right) $ in equation (\ref{eq:PE_Probit}).\ Similarly to the
IV-Tobit model, the IV-Probit model can be estimated by MLE or by the
two-step approach identical to the one described in Section~\ref{sec:Model},
except the second step uses the standard Probit estimator in place of the
Tobit estimator. Then the bounds on $PE_{j}^{\Pr }\left( h\right) $ are
estimated as in equation (\ref{eq:bounds_hat_PE_generic}).%
{}
Confidence intervals for $PE_{j}^{\Pr }(h)$ can be computed exactly as
described above.

\bigskip

\section{Average Partial Effects and Other Counterfactuals}\label%
{sec:APEs}

In addition to the partial effects at a given $h$, researchers are often
interested in the Average Partial Effects%
\begin{equation}
APE_{j}^{\text{Tob}}\equiv \mathrm{E}\left[ PE_{j}^{\text{Tob}}\left(
H_{i}^{\ast }\right) \right] \text{ and }APE_{j}^{\Pr }\equiv \mathrm{E}%
\left[ PE_{j}^{\Pr }\left( H_{i}^{\ast }\right) \right] ,
\label{eq:APEs def}
\end{equation}%
which are the partial effects $PE_{j}\left( h\right) $ averaged with respect
to the distribution of {}$H_{i}^{\ast }=\left(
X_{i}^{\ast },W_{i}^{\prime }\right) ^{\prime }$. Define%
\begin{equation*}
APE_{j}^{\text{Tob}}\left( \sigma _{U^{\ast }}^{2}\right) \equiv \mathrm{E}%
\left[ PE_{j}^{\text{Tob}}\left( H_{i}^{\ast },\sigma _{U^{\ast
}}^{2}\right) \right] \text{ and }APE_{j}^{\Pr }\left( \sigma _{U^{\ast
}}^{2}\right) \equiv \mathrm{E}\left[ PE_{j}^{\Pr }\left( H_{i}^{\ast
},\sigma _{U^{\ast }}^{2}\right) \right] .
\end{equation*}

Note that the distribution of $X_{i}^{\ast }$ is not directly observable due
to the EiV. Averaging $PE_{j}\left( h\right) $ with respect to the
distribution of the observed $H_{i}=\left( X_{i},W_{i}^{\prime }\right)
^{\prime }$ would result in biased estimators of the APEs. To account for
this, in the Appendix we show that these APEs can be calculated as%
\begin{align}
APE_{j}^{\text{Tob}}\left( \sigma _{U^{\ast }}^{2}\right) & =\mathrm{E}\left[
\Phi \left( \frac{\theta _{01}\pi _{01}^{\prime }Z_{i}+\left( \theta
_{01}\pi _{02}+\theta _{02}\right) ^{\prime }W_{i}}{\sqrt{2\sigma _{U^{\ast
}}^{2}-\sigma _{U}^{2}+\theta _{01}^{2}\sigma _{V}^{2}}}\right) \right]
\theta _{0j},  \label{eq:true APE Tobit} \\
APE_{j}^{\Pr }\left( \sigma _{U^{\ast }}^{2}\right) & =\mathrm{E}\left[ \phi
\left( \frac{\theta _{01}\pi _{01}^{\prime }Z_{i}+\left( \theta _{01}\pi
_{02}+\theta _{02}\right) ^{\prime }W_{i}}{\sqrt{2\sigma _{U^{\ast
}}^{2}-\sigma _{U}^{2}+\theta _{01}^{2}\sigma _{V}^{2}}}\right) \right] 
\frac{\theta _{0j}}{\sqrt{2\sigma _{U^{\ast }}^{2}-\sigma _{U}^{2}+\theta
_{01}^{2}\sigma _{V}^{2}}}.  \label{eq:true APE Probit}
\end{align}%
Hence, for any given value of $\sigma _{U^{\ast }}^{2}$, these APEs can be
estimated by%
\begin{align*}
\widehat{APE}_{j}^{\text{Tob}}\left( \sigma _{U^{\ast }}^{2}\right) & =\frac{%
1}{n}\sum_{i=1}^{n}\Phi \left( \frac{\hat{\theta}_{1}\hat{\pi}_{1}^{\prime
}Z_{i}+\left( \hat{\theta}_{1}\hat{\pi}_{2}+\hat{\theta}_{2}\right) ^{\prime
}W_{i}}{\sqrt{2\sigma _{U^{\ast }}^{2}-\hat{\sigma}_{U}^{2}+\hat{\theta}%
_{1}^{2}\hat{\sigma}_{V}^{2}}}\right) \hat{\theta}_{j}, \\
\widehat{APE}_{j}^{\Pr }\left( \sigma _{U^{\ast }}^{2}\right) & =\frac{1}{n}%
\sum_{i=1}^{n}\phi \left( \frac{\hat{\theta}_{1}\hat{\pi}_{1}^{\prime
}Z_{i}+\left( \hat{\theta}_{1}\hat{\pi}_{2}+\hat{\theta}_{2}\right) ^{\prime
}W_{i}}{\sqrt{2\sigma _{U^{\ast }}^{2}-\hat{\sigma}_{U}^{2}+\hat{\theta}%
_{1}^{2}\hat{\sigma}_{V}^{2}}}\right) \frac{\hat{\theta}_{j}}{\sqrt{2\sigma
_{U^{\ast }}^{2}-\hat{\sigma}_{U}^{2}+\hat{\theta}_{1}^{2}\hat{\sigma}%
_{V}^{2}}}.
\end{align*}%
Finally, the estimated bounds on the APEs are obtained by finding the
minimum and maximum over $\sigma _{U^{\ast }}^{2}\in \left[ \hat{\underline{%
\sigma }}_{U^{\ast }}^{2},\hat{\sigma}_{U}^{2}\right] $. These can be easily
computed numerically, since $\widehat{APE}_{j}\left( v\right) $ are smooth
functions of a scalar argument $v$. Our two-step approach to inference also
applies to the APEs with a minimal modification. The only difference is that
in Step 2 the construction of the standard error $s_{\widehat{APE}%
_{j}}\left( \sigma _{U^{\ast }}^{2}\right) $ as usual needs to account for
the sampling variability in both the parameter estimators and the data
entering the expressions for the APEs directly.

It is also straightforward to apply our analysis to other counterfactuals,
including partial effects and APEs of discrete covariates, as well as to the
ordered Probit and two-sided Tobit models. Proposition~\ref{prop:Ustar
bounds} and the bounds on $\sigma _{U^{\ast }}^{2}$ remain the same, and
hence the estimation and inference procedures remain unchanged, except for
different formulas in equations~(\ref{eq:PE_ProbitTobit_sigStar})-(\ref%
{eq:bounds_hat_PE_generic}) corresponding to the counterfactuals of interest.

\bigskip

\section{Numerical Illustration}\label{sec:MC}

We simulate a Tobit model with endogenous and mismeasured $X_{i}^{\ast }$,
as in equations (\ref{eq:dgp Y true})-(\ref{eq:ME}), with $W_{i}=1$, $%
Z_{i}\sim N\left( 0,1\right) $, $\left( \theta _{01},\theta _{02},\sigma
_{V^{\ast }},\sigma _{U^{\ast }},\sigma _{\varepsilon },\pi _{01},\pi
_{02}\right) =\left( 2,1,1,1,1,1,0\right) $, and $n=1000$. {}Figure~\ref%
{fig:MC} plots the results for the Partial Effects (PEs) of $X_{i}^{\ast }$
at the population mean values of the covariates.

We consider a range of designs corresponding to the true values of $\rho
_{U^{\ast }V^{\ast }}\in \left[ -0.95,0.95\right] $ on the horizontal axis.
For each $\rho _{U^{\ast }V^{\ast }}$, the figure shows the true PE
(\textquotedblleft true\textquotedblright ), the true (population) bounds
for the PE obtained using Proposition~\ref{prop:Ustar bounds}
(\textquotedblleft true bounds\textquotedblright ), as well as the medians
over the Monte Carlo replications of the estimated lower and upper bounds on
the PE (\textquotedblleft LB\textquotedblright\ and \textquotedblleft
UB\textquotedblright ) and the corresponding $95\%$ confidence intervals
(\textquotedblleft CI\textquotedblright ) based on the two-step IV-Tobit
estimator. The true bounds for the PE are calculated using the point
identified parameters $\theta _{0}$, $\sigma _{U}^{2}$, $\sigma _{UV}$, and $%
\sigma _{V}^{2}$, see equations (\ref{eq:dgp Y wo stars gen})-(\ref{eq:dgp
UV wo stars gen}). For comparison, we also include the results for the PE
calculated using the standard naive IV-Tobit estimator (\textquotedblleft
naive\textquotedblright ) and the corresponding confidence intervals
(\textquotedblleft CI naive\textquotedblright ). The \textquotedblleft
naive\textquotedblright\ estimators of the partial effects are $\widehat{PE}%
_{1}^{\text{Tob}}\left( h,\widehat{\sigma }_{U}^{2}\right) $ and $\widehat{PE%
}_{1}^{\Pr }\left( h,\widehat{\sigma }_{U}^{2}\right) $, i.e., they replace $%
\sigma _{U^{\ast }}^{2}$ with $\widehat{\sigma }_{U}^{2}$ in equation (\ref%
{eq:hat PE-ProbitTobit(sigUs2)}).

\begin{figure}[!h]
\centering%
\includegraphics[trim = {0.5cm 0.7cm 0.5cm
0.2cm},clip,width=\textwidth,keepaspectratio]{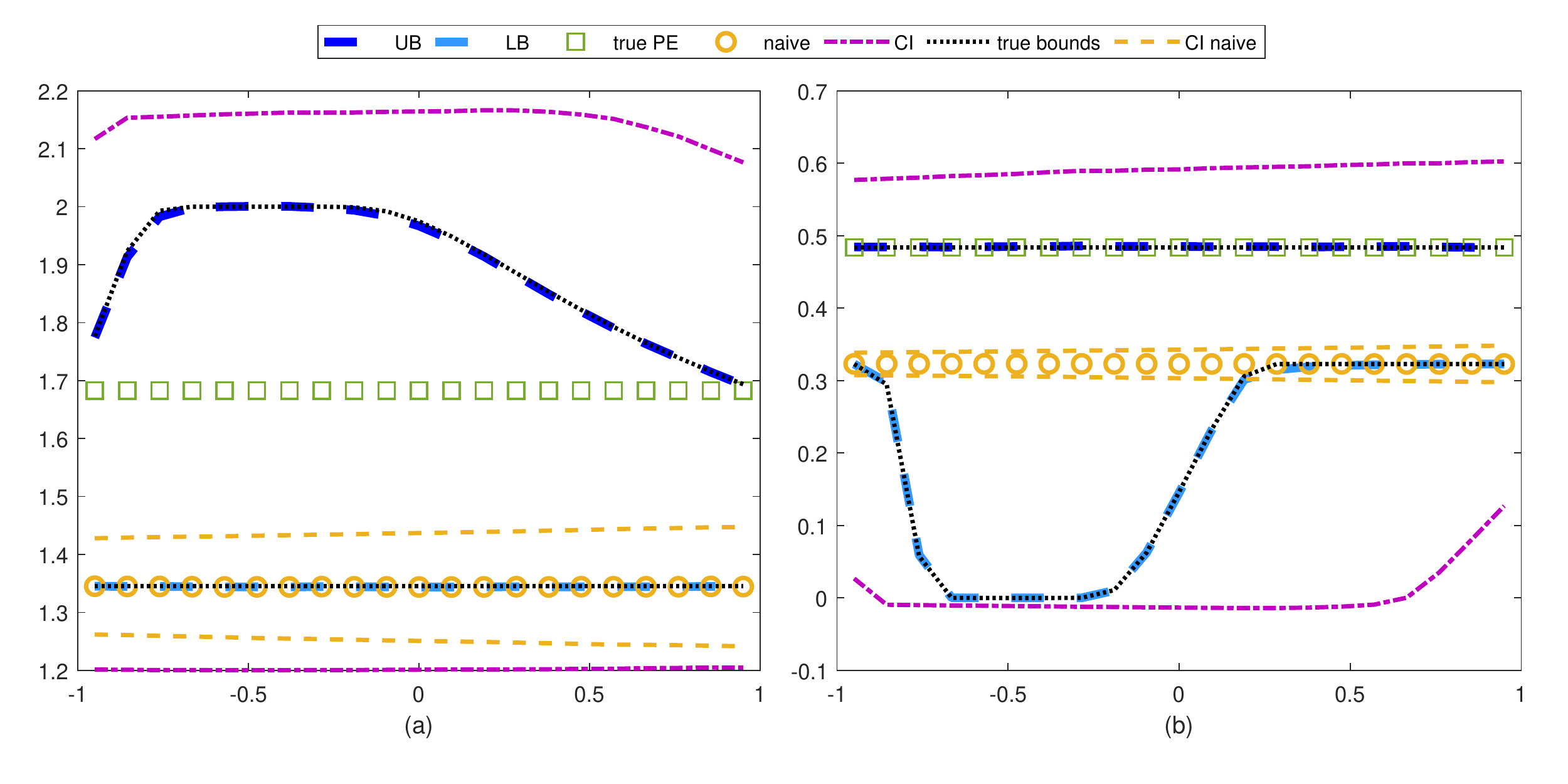}
\caption{Simulation results for partial effects on: (a) expectations, $%
PE_{1}^{\text{Tob}}\left( h\right)$, and (b) probability, $PE_{1}^{\text{%
Pr}}\left( h\right) $. Values of $\protect\rho_{U^{\ast }V^{\ast }}$ are on
the horizontal axis.}
\label{fig:MC}
\end{figure}

Figure~\ref{fig:MC} (a) shows the bounds on $PE_{h_{1}}^{\text{Tob}}\left(
h\right) $, while Figure~\ref{fig:MC} (b) considers $PE_{1}^{\text{Pr}%
}\left( h\right) $. As expected, the true PE is between the lower and upper
bounds for all values of $\rho _{U^{\ast }V^{\ast }}$. By construction, the
\textquotedblleft naive\textquotedblright\ \ IV-Tobit estimator of $PE_{1}^{%
\text{Tob}}\left( h\right) $ coincides with one of the bounds. In both
panels, the \textquotedblleft naive\textquotedblright\ estimates are below
the true values for every $\rho _{U^{\ast }V^{\ast }}$, and the
\textquotedblleft naive\textquotedblright\ IV-Tobit confidence intervals do
not include the true PE. In this design, the identified set and the
confidence intervals for the true partial effects are much wider than those
of the \textquotedblleft naive\textquotedblright\ estimator. The relative
width of the identified set and the confidence intervals depends on the
specific parameter values. In contrast to these simulations, in the example
of the next section, the identified set is very narrow and the confidence
intervals for the partial effects have width similar to those of the naive
estimator.

To gain some intuition about the shape of the bounds in Figure~\ref{fig:MC},
notice that when $\theta _{01}>0$, measurement error in $X_{i}^{\ast }$
introduces a negative correlation between $X_{i}$ and $U_{i}$. Thus,
observing $\rho _{UV}>0$ is only consistent with $\rho _{U^{\ast }V^{\ast
}}>0$, but not with $\rho _{U^{\ast }V^{\ast }}\leq 0$, i.e., implies
positive correlation due to the structural endogeneity. On the other hand,
observing $\rho _{UV}<0$ can be explained both by the effect of EiV combined
with $\rho _{U^{\ast }V^{\ast }}\gtrless 0$, and by $\rho _{U^{\ast }V^{\ast
}}<0$ without any EiV. Thus, when $\rho _{UV}<0$ it is harder to disentangle
structural endogeneity and EiV. As a consequence, the true and the estimated
correct bounds on the PEs in Figure~\ref{fig:MC} are wider for the negative
values of $\rho _{U^{\ast }V^{\ast }}$.

\bigskip

\section{Empirical Application}\label{sec:Empirical}

We illustrate the proposed methods in the classic application estimating the Tobit and Probit
models for women's labor force participation. Using NLSY97 we construct an up-to-date dataset similar to the well known dataset of \cite{Mroz1987Ecta}, who estimates Tobit and related models to explain
married women's hours of work. The data contains 1138 women continuously married in 2018, 79\% of which
report working non-zero hours. For Tobit, the dependent variable is the
number of hours worked (\textit{hours}), and for Probit, the dependent
variable is working at some point during the year ($1\{hours>0\}$). In both
models, the covariates are age, education, experience, experience squared,
nonwife income in thousands (\textit{nwifeinc}), number of children less
than six years of age, number of children between 6 and 18 years of age, race-ethnicity indicator variables, 
and an intercept. Nonwife income can be correlated with the unobserved
characteristics (structural endogeneity), and income variables are also
known to be frequently mismeasured. Spouse's education (\textit{speduc}) is used as an
instrument for \textit{nwifeinc}. This specification is used in, e.g., \cite%
{Wooldridge2010Book}. 

\begin{table}[!h]
\rowcolors{1}{white}{myEvenRowsColor}
  \begin{center}
  \begin{tabular}{|c|c|cc|cc|}
    \hline\hline
   &  Tobit & IV-Tobit & CI for IV-Tobit & [LB, UB] for APE & CI for APE\\ \hline
    nwifeinc &  $-1.36$ & $-4.07$ & $[-7.06,-1.08]$ & $[-4.12,-4.07]$ & $[-7.39,-0.928]$\\ 
    educ &  $29.8$ & $40.4$ & $[18.9,61.9]$ & $[40.4,40.9]$ & $[17.6,64.5]$\\ 
    exper &  $200$ & $207$ & $[154,260]$ & $[207,210]$ & $[148,268]$\\ 
    exper2 &  $-2.22$ & $-2.81$ & $[-4.95,-0.671]$ & $[-2.85,-2.81]$ & $[-5.15,-0.592]$\\ 
    age &  $-91.4$ & $-76.9$ & $[-117,-36.6]$ & $[-77.9,-76.9]$ & $[-121,-35.3]$\\ 
  \hline\hline
  \end{tabular}
  \caption{Tobit. Average Partial Effects on Expectation. $n=1138$.}
  \label{table:Emp-Tobit-PEM-E-w-CI}
  \end{center}
\end{table}

\begin{table}[!h]
\rowcolors{1}{white}{myEvenRowsColor}
  \begin{center}
  \begin{tabular}{|c|c|cc|cc|}
    \hline\hline
   &  Tobit & IV-Tobit & CI for IV-Tobit & [LB, UB] for APE & CI for APE\\ \hline
    nwifeinc &  $-0.024$ & $-0.075$ & $[-0.134,-0.015]$ & $[-0.075,-0.072]$ & $[-0.138,-0.015]$\\ 
    educ &  $0.528$ & $0.741$ & $[0.313,1.17]$ & $[0.718,0.741]$ & $[0.290,1.21]$\\ 
    exper &  $3.54$ & $3.80$ & $[2.90,4.69]$ & $[3.67,3.80]$ & $[2.79,4.89]$\\ 
    exper2 &  $-0.039$ & $-0.052$ & $[-0.090,-0.013]$ & $[-0.052,-0.050]$ & $[-0.094,-0.013]$\\ 
    age &  $-1.62$ & $-1.41$ & $[-2.15,-0.669]$ & $[-1.41,-1.37]$ & $[-2.27,-0.563]$\\ 
  \hline\hline
  \end{tabular}
  \caption{Tobit. Average Partial Effects on Probability. All numbers are multiplied by 100. $n=1138$.}
  \label{table:Emp-Tobit-PEM-P-w-CI}
  \end{center}
\end{table}

\begin{table}[!h]
\rowcolors{1}{white}{myEvenRowsColor}
  \begin{center}
  \begin{tabular}{|c|c|cc|cc|}
    \hline\hline
   &  Probit & IV-Probit & CI for IV-Probit & [LB, UB] for APE & CI for APE\\ \hline
    nwifeinc &  $-0.028$ & $-0.078$ & $[-0.174,0.019]$ & $[-0.079,-0.078]$ & $[-0.188,0.024]$\\ 
    educ &  $0.382$ & $0.594$ & $[-0.113,1.30]$ & $[0.594,0.606]$ & $[-0.143,1.40]$\\ 
    exper &  $-0.094$ & $0.091$ & $[-1.27,1.45]$ & $[0.091,0.093]$ & $[-1.37,1.56]$\\ 
    exper2 &  $0.170$ & $0.156$ & $[0.084,0.228]$ & $[0.156,0.159]$ & $[0.082,0.247]$\\ 
    age &  $-2.15$ & $-1.87$ & $[-3.15,-0.587]$ & $[-1.91,-1.87]$ & $[-3.34,-0.559]$\\ 
  \hline\hline
  \end{tabular}
  \caption{Probit. Average Partial Effects on Probability. All numbers are multiplied by 100. $n=1138$.}
  \label{table:Emp-Probit-PEM-P-w-CI}
  \end{center}
\end{table}

Tables \ref{table:Emp-Tobit-PEM-E-w-CI}-\ref{table:Emp-Tobit-PEM-P-w-CI}
contain the results for average partial effects for Tobit. Table \ref%
{table:Emp-Tobit-PEM-E-w-CI} contains the results on average partial effects on
expectation, $APE^{\text{Tob}}$, while Table \ref{table:Emp-Tobit-PEM-P-w-CI}
contains average partial effects on probability, $APE^{\text{Pr}}$.  
 In both tables, the
first column (\textquotedblleft Tobit") provides the average partial effects for
different covariates estimated by the standard Tobit MLE, where all covariates are
assumed to be exogenous. 
The remaining columns are based on the two-step IV-Tobit estimator, where 
\textit{speduc} is used as an instrument for the endogenous \textit{nwifeinc.}
The second column (\textquotedblleft IV-Tobit") contains the naive
estimators of the average partial effects, followed by the $95\%$ confidence
intervals (column \textquotedblleft CI for IV-Tobit"). Column
``[LB, UB] for APE'' provides the proposed estimated bounds
for the average partial effects that account for both types of endogeneity. The last
column contains the corresponding confidence intervals for the average partial
effects.

In both Tables \ref{table:Emp-Tobit-PEM-E-w-CI} and \ref%
{table:Emp-Tobit-PEM-P-w-CI}, we observe that the confidence intervals for
the correct average partial effects are only slightly wider than the
naive ones of the IV-Tobit. In particular, using the correct inference approach
does not change any of the conclusions about the effects of the variables
being statistically significant.

Table \ref{table:Emp-Probit-PEM-P-w-CI} contains the corresponding results
for Probit. Again, the confidence intervals for the correct partial effects
are not much wider than the naive ones, and as before,
using the correct inference approach does not change any of the conclusions about the
effects of the variables being statistically significant.

Overall, we find that the proposed valid confidence intervals for the APEs are only slightly wider than those obtained from the naive (and generally invalid) IV-Tobit/IV-Probit estimators. The point estimates of the valid bounds in columns ``[LB, UB] for APE'' are narrow. We interpret these results as (strong) evidence in favor of practical usefulness of our method. Practitioners often worry that methods providing partial identification of parameters of interest achieve robustness at the cost of producing bounds that are too wide to be useful in practice. Our empirical application shows that this concern need not be an issue for our method.

\bigskip

\section{Relaxing Distributional Assumptions}\label{sec:Relax Gauss}

We have considered the classic IV-Tobit and IV-Probit settings, which
require $V_{i}^{\ast }$ and $\varepsilon _{i}$ to be Gaussian in order for $%
V_{i}$ and $U_{i}$ to be Gaussian, as in \cite{SmithBlundell1986Ecta}, \cite%
{RiversVuong1988JoE}, and \cite{Wooldridge2010Book}. However, at least in
the model without EiV, the control variable approach does not require $%
V_{i}^{\ast }$ to be Gaussian. Likewise, we would like to relax the
assumption of Gaussianity on $\varepsilon _{i}$. In this section we propose
an approach that weakens the distributional assumptions, while still
providing a simple method for computing the identified set for the partial
effects of interest. We focus on the IV-Tobit settings.

We model the joint distribution of $U_{i}^{\ast }$ and $V_{i}^{\ast }$ as a
mixture of $J$ bivariate normals. Specifically, the joint p.d.f. of $U^{\ast
}$ and $V^{\ast }$ takes the form %
\begin{equation*}
f_{U^{\ast }V^{\ast }}(u,v)=\sum_{j=1}^{J}p_{V^{\ast },j}\phi (u,v;\mu
_{U^{\ast }V^{\ast },j},\Sigma _{U^{\ast }V^{\ast },j}),
\end{equation*}%
where $p_{V^{\ast },j}>0$ are the mixing weights, and 
\begin{equation*}
\mu _{U^{\ast }V^{\ast },j}=%
\begin{pmatrix}
0 \\ 
\mu _{V^{\ast },j}%
\end{pmatrix}%
,\quad \Sigma _{U^{\ast }V^{\ast },j}=%
\begin{pmatrix}
\sigma _{U^{\ast }}^{2} & \sigma _{U^{\ast }V^{\ast },j} \\ 
\sigma _{U^{\ast }V^{\ast },j} & \sigma _{V^{\ast },j}^{2}%
\end{pmatrix}%
,
\end{equation*}%
and $\phi (\cdot ,\cdot ;\mu ,\Sigma )$ stands for the p.d.f. of a bivariate
normal with mean $\mu $ and variance-covariance matrix $\Sigma $. Notice
that in this parameterization the marginal distribution of $U_{i}^{\ast }$
is $N(0,\sigma _{U^{\ast }}^{2})$ as in the standard Tobit model, whereas
the marginal distribution of $V_{i}^{\ast }$ is flexibly modelled as a
mixture of $J$ normals. Since $\sigma _{U^{\ast }V^{\ast },j}$ can vary over 
$j$, this model also allows rich patterns of dependency between $U_{i}^{\ast
}$ and $V_{i}^{\ast }$. Similarly, we model the distribution of $\varepsilon
_{i}$ by a mixture of $L$ normals:%
\begin{equation*}
f_{\varepsilon }(\varepsilon )=\sum_{\ell =1}^{L}p_{\varepsilon ,\ell }\phi
(\varepsilon ;\mu _{\varepsilon ,\ell },\sigma _{\varepsilon ,\ell }^{2}),
\end{equation*}%
where $p_{\varepsilon ,\ell }>0$ are the mixing weights, and $\phi (\cdot
;\mu ,\sigma ^{2})$ stands for the p.d.f. of a $N\left( \mu ,\sigma
^{2}\right) $ distribution, and we denote the corresponding c.d.f. by $\Phi
(\cdot ;\mu ,\sigma ^{2})$.

Since $\varepsilon _{i}$ is independent from $\left( U_{i}^{\ast
},V_{i}^{\ast }\right) $, the joint distribution of $\left(
U_{i},V_{i}\right) =\left( U_{i}^{\ast }-\theta _{01}\varepsilon
_{i},V_{i}^{\ast }+\varepsilon _{i}\right) $ is a mixture of $J\times L$
bivariate normals, and its p.d.f. is given by 
\begin{gather*}
f_{UV}(u,v)=\sum_{j=1}^{J}\sum_{\ell =1}^{L}p_{V^{\ast },j}p_{\varepsilon
,\ell }\phi (u,v;\mu _{UV,j\ell },\Sigma _{UV,j\ell }),\text{ where} \\
\mu _{UV,j\ell }=%
\begin{pmatrix}
\mu _{U,\ell } \\ 
\mu _{V,j\ell }%
\end{pmatrix}%
=%
\begin{pmatrix}
-\theta _{01}\mu _{\varepsilon ,\ell } \\ 
\mu _{V^{\ast },j}+\mu _{\varepsilon ,\ell }%
\end{pmatrix}%
,\quad \Sigma _{UV,j\ell }=%
\begin{pmatrix}
\sigma _{U,j\ell }^{2} & \sigma _{UV,j\ell } \\ 
\sigma _{UV,j\ell } & \sigma _{V,j\ell }^{2}%
\end{pmatrix}%
,
\end{gather*}%
\begin{equation}
\sigma _{U,j\ell }^{2}=\sigma _{U^{\ast }}^{2}+\theta _{01}^{2}\sigma
_{\varepsilon ,\ell }^{2},\quad \sigma _{V,j\ell }^{2}=\sigma _{V^{\ast
},j}^{2}+\sigma _{\varepsilon ,\ell }^{2},\quad \sigma _{UV,j\ell }=\sigma
_{U^{\ast }V^{\ast },j}-\theta _{01}\sigma _{\varepsilon ,\ell }^{2}.
\label{eq:general reduced form to structural relationship}
\end{equation}%
We impose standard constraints $\sum_{j=1}^{J}p_{V^{\ast },j}=1$ and $%
\sum_{\ell =1}^{L}p_{\varepsilon ,\ell }=1$. In addition, we need location
normalizations on the distributions of $V_{i}^{\ast }$ and $\varepsilon _{i}$%
. We follow the standard approach and assume that $E\left[ V_{i}^{\ast }%
\right] =0$ and $E\left[ \varepsilon _{i}\right] =0$, which are imposed by
the restrictions $\sum_{j=1}^{J}p_{V^{\ast },j}\mu _{V^{\ast },j}=0$ and $%
\sum_{\ell =1}^{L}p_{\varepsilon ,\ell }\mu _{\varepsilon ,\ell }=0$.
Alternatively, it is possible to normalize the medians instead of the means.
For example, the restriction $\sum_{\ell =1}^{L}p_{\varepsilon ,\ell }\Phi
(0;\mu _{\varepsilon ,\ell },\sigma _{\varepsilon ,\ell }^{2})=1/2$ imposes
normalization $\limfunc{median}\left( \varepsilon _{i}\right) =0$, which in
particular allows $\varepsilon _{i}$ to be a non-classical measurement error.%
\footnote{%
It is also possible to impose a normalization on the modes of the
distributions of $\varepsilon _{i}$ and/or $V_{i}^{\ast }$, although this is
less convenient computationally.} {}

This model naturally generalizes the classic Gaussian model considered in
the previous sections. Taking $J=1$ restricts $V_{i}^{\ast }$ to be
Gaussian. Taking $L=1$ corresponds to $\varepsilon _{i}$ being Gaussian.

Thus, we consider the IV-Mixed-Tobit model that consists of equations~(\ref%
{eq:dgp Y wo stars gen})-(\ref{eq:dgp X wo stars gen}) and replaces equation
(\ref{eq:dgp UV wo stars gen}) with the assumption that%
\begin{equation}
f_{UV}(u,v)=\sum_{k=1}^{K}p_{k}\phi (u,v;\mu _{UV,k},\Sigma _{UV,k}),\qquad
\mu _{UV,k}=%
\begin{pmatrix}
\mu _{U,k} \\ 
\mu _{V,k}%
\end{pmatrix}%
,\quad \Sigma _{UV,k}=%
\begin{pmatrix}
\sigma _{U,k}^{2} & \sigma _{UV,k} \\ 
\sigma _{UV,k} & \sigma _{V,k}^{2}%
\end{pmatrix}%
,  \label{eq:f_UV}
\end{equation}%
where $\sum_{k=1}^{K}p_{k}\mu _{UV,k}=\left( 0,0\right) ^{\prime }$ and $%
\sum_{k=1}^{K}p_{k}=1$.

First, we discuss identification of the parameters of the above
IV-Mixed-Tobit model. 
As in Section~\ref{sec:Analysis}, $\pi _{0}$ is immediately identified by
the first stage regression of $X_{i}$ on $Z_{i}$ and $W_{i}$. Identification
of the remaining parameters is less straightforward and is established by
the following proposition. This identification result appears to be new.

\begin{proposition}
\label{prop:IV-Mixed-Tobit ID} Suppose $\pi _{01}\neq 0$ and $%
E[(Z_{i}^{\prime },W_{i}^{\prime })^{\prime }(Z_{i}^{\prime },W_{i}^{\prime
})]$ has full rank. Then parameters $\theta _{0}$, $\pi _{0}$, and $\left\{
p_{k},\mu _{UV,k},\Sigma _{UV,k}\right\} _{k=1}^{K}$ are identified (up to
relabelling of the mixture components indexed by $k$).
\end{proposition}

Proposition~\ref{prop:IV-Mixed-Tobit ID} establishes identification of
parameters $\theta _{0},\pi _{0},\left\{ p_{k},\mu _{UV,k},\Sigma
_{UV,k}\right\} _{k=1}^{K}$. These parameters can be estimated by the MLE. The log-likelihood can be optimized using the EM algorithm commonly employed for estimation of mixture models. Alternatively, for moderate $K$, it is also feasible to optimize the log-likelihood directly since both the log-likelihood and its Jacobian are available in closed form.
From a practical perspective, we do not recommend models with large $K$,
because Gaussian mixture models are known to be highly flexible even with
relatively small $K$. 

Our goal is to provide bounds on the partial effects. To do this, as in
Section~\ref{sec:Analysis}, the key is to construct the identified set for $%
\sigma _{U^{\ast }}^{2}$. Notice that, for every $k$, $\Sigma _{UV,k}$ is
equal to $\Sigma _{UV,j\ell }$ for some $j$ and $\ell $ satisfying the
restrictions given by equation 
\eqref{eq:general reduced form to structural
relationship}. Since these restrictions have the same structure as the ones
in equation \eqref{eq:sigma2 w and wo stars}, we can apply the result of
Proposition~\ref{prop:Ustar bounds} with a given $\Sigma _{UV,k}$ to
construct an identified set for $\sigma _{U^{\ast }}^{2}$ given by 
\begin{equation*}
\mathcal{I}_{k}=\left[ \underline{\sigma }_{U^{\ast },k}^{2},\sigma
_{U,k}^{2}\right] ,
\end{equation*}%
where $\underline{\sigma }_{U^{\ast },k}^{2}$ is computed as in equation~(%
\ref{eq:bound s2Us alt}) with $\sigma _{U,k}^{2}$, $\sigma _{V,k}^{2}$ and $%
\sigma _{UV,k}$ in place of $\sigma _{U}^{2}$, $\sigma _{V}^{2}$, and $%
\sigma _{UV}$, respectively. Hence, we can construct an identified set for $%
\sigma _{U^{\ast }}^{2}$ by intersecting $\mathcal{I}_{k}$ for $k\in
\{1,\ldots ,K\}$, i.e., the bounds for $\sigma _{U^{\ast }}^{2}$ can be
constructed as 
\begin{equation}
\sigma _{U^{\ast }}^{2}\in \mathcal{I}\equiv \bigcap_{k}\mathcal{I}_{k}=%
\left[ \max_{k}\underline{\sigma }_{U^{\ast },k}^{2},\min_{k}\sigma
_{U,k}^{2}\right] .  \label{eq:general bounds for sigma U star}
\end{equation}

\begin{proposition}
\label{prop:general Ustar bounds} The identified set for $\sigma _{U^{\ast
}}^{2}$ given by equation~(\ref{eq:general bounds for sigma U star}) is
sharp.
\end{proposition}

Once the identified set for $\sigma _{U^{\ast }}^{2}$ is constructed, the
rest of the analysis follows as in the Gaussian model in Sections~\ref%
{sec:Model}-\ref{sec:Analysis}. In particular, sharp bounds for the partial
effects are constructed as in equation \eqref{eq:bounds_PE_generic} with $%
\mathcal{I}$ as the identified set for $\sigma _{U^{\ast }}^{2}$. That is,
the lower and upper bounds for partial effects for the $j^{th}$ covariate, $%
PE_{j}\left( h\right) $, are given by 
\begin{equation*}
\min_{\sigma _{U^{\ast }}^{2}\in \mathcal{I}}PE_{j}\left( h,\sigma _{U^{\ast
}}^{2}\right) \text{\quad and\quad }\max_{\sigma _{U^{\ast }}^{2}\in 
\mathcal{I}}PE_{j}\left( h,\sigma _{U^{\ast }}^{2}\right) .
\end{equation*}%
Note that set $\mathcal{I}$ is a closed interval, as in the original
IV-Tobit model. Thus, the discussion concerning the implementation of the
identified sets for the partial effects that follows equation~(\ref%
{eq:bounds_PE_generic}) also applies to the IV-Mixed-Tobit model of this
section. {}

\bigskip

\section{Conclusion}

Both structural endogeneity and mismeasurement of covariates are pervasive
in economic data. Estimating partial effects and other counterfactuals using
such data is an important practical problem. This paper addresses the
problem in the classic Probit and Tobit models. The relative simplicity of
these nonlinear models allows for transparent analysis and practical
solutions. The paper provides simple estimators and confidence intervals
that are easy to implement. The paper also shows how the proposed solutions
can be extended to settings with non-Gaussian unobservables.

\bigskip

\section*{Acknowledgements}
Evdokimov gratefully acknowledges the support from the Spanish MICIU/AEI via grants RYC2020-030623-I, funded by MICIU/AEI/10.13039/501100011033 and by ``ESF Investing in your future'', PID2019-107352GB-I00, PID2022-140825NB-I00, and Severo Ochoa Programme CEX2019-000915-S. Zeleneev gratefully acknowledges the generous funding from the UK Research and Innovation (UKRI) under the UK government’s Horizon Europe funding
guarantee (Grant Ref: EP/X02931X/1).

\bigskip

\nocite{NLSY97}
\bibliographystyle{ecta}
\bibliography{ME_Probit_out}

\bigskip

 \begin{appendices}%

\numberwithin{equation}{section}

\section{Appendix\label{sec:Appx}}

\medskip

\subsection{Proof of Proposition~\protect\ref{prop:Ustar bounds}}

Note that $\sigma _{U^{\ast }V^{\ast }}^{2}\leq \sigma _{U^{\ast
}}^{2}\sigma _{V^{\ast }}^{2}$ combined with equation~(\ref{eq:sigma2 w and
wo stars}) implies%
\begin{eqnarray}
0 &\leq &\sigma _{U^{\ast }}^{2}\sigma _{V^{\ast }}^{2}-\sigma _{U^{\ast
}V^{\ast }}^{2}=\left( \sigma _{U}^{2}-\theta _{01}^{2}\sigma _{\varepsilon
}^{2}\right) \left( \sigma _{V}^{2}-\sigma _{\varepsilon }^{2}\right)
-\left( \sigma _{UV}+\theta _{01}\sigma _{\varepsilon }^{2}\right) ^{2}%
{} 
\notag \\
&=&\sigma _{U}^{2}\sigma _{V}^{2}-\sigma _{UV}^{2}-\sigma _{\varepsilon
}^{2}\left( \sigma _{V}^{2}\theta _{01}^{2}+2\sigma _{UV}\theta _{01}+\sigma
_{U}^{2}\right) ,\text{ and hence}  \notag \\
0 &\leq &\sigma _{\varepsilon }^{2}\leq \frac{\sigma _{U}^{2}\sigma
_{V}^{2}-\sigma _{UV}^{2}}{\sigma _{V}^{2}\theta _{01}^{2}+2\sigma
_{UV}\theta _{01}+\sigma _{U}^{2}}.  \label{eq:sigma2 eps CS}
\end{eqnarray}%
Since $\left\vert \rho _{UV}\right\vert <1$, the denominator in this
fraction is positive. Since $\sigma _{\varepsilon }^{2}\leq \sigma _{V}^{2}$%
, let%
\begin{equation*}
\overline{\sigma }_{\varepsilon }^{2}\equiv \min \left\{ \frac{\sigma
_{U}^{2}\sigma _{V}^{2}-\sigma _{UV}^{2}}{\sigma _{V}^{2}\theta
_{01}^{2}+2\sigma _{UV}\theta _{01}+\sigma _{U}^{2}},\sigma _{V}^{2}\right\}
.
\end{equation*}%
Then%
\begin{equation}
\underline{\sigma }_{U^{\ast }}^{2}\leq \sigma _{U^{\ast }}^{2}\leq \sigma
_{U}^{2},\qquad \underline{\sigma }_{U^{\ast }}^{2}\equiv \max \left\{
\sigma _{U}^{2}-\theta _{01}^{2}\overline{\sigma }_{\varepsilon
}^{2},0\right\} .  \label{eq:bound s2Us}
\end{equation}%
Note that%
\begin{equation*}
\sigma _{U}^{2}-\theta _{01}^{2}\frac{\sigma _{U}^{2}\sigma _{V}^{2}-\sigma
_{UV}^{2}}{\sigma _{V}^{2}\theta _{01}^{2}+2\sigma _{UV}\theta _{01}+\sigma
_{U}^{2}}=\frac{\sigma _{U}^{4}+2\sigma _{U}^{2}\sigma _{UV}\theta
_{01}+\theta _{01}^{2}\sigma _{UV}^{2}}{\sigma _{V}^{2}\theta
_{01}^{2}+2\sigma _{UV}\theta _{01}+\sigma _{U}^{2}}=\frac{\left( \theta
_{01}\sigma _{UV}+\sigma _{U}^{2}\right) ^{2}}{\sigma _{V}^{2}\theta
_{01}^{2}+2\sigma _{UV}\theta _{01}+\sigma _{U}^{2}}.
\end{equation*}%
Thus, $\underline{\sigma }_{U^{\ast }}^{2}$ in equation~(\ref{eq:bound s2Us}%
) can be equivalently written as%
\begin{equation*}
\underline{\sigma }_{U^{\ast }}^{2}\equiv \max \left\{ \frac{\left( \theta
_{01}\sigma _{UV}+\sigma _{U}^{2}\right) ^{2}}{\sigma _{V}^{2}\theta
_{01}^{2}+2\sigma _{UV}\theta _{01}+\sigma _{U}^{2}},\sigma _{U}^{2}-\theta
_{01}^{2}\sigma _{V}^{2}\right\} ,
\end{equation*}%
where the fraction is always non-negative, ensuring that $\underline{\sigma }%
_{U^{\ast }}^{2}\geq 0$. Similarly, equation~(\ref{eq:sigma2 w and wo stars}%
) implies that $\sigma _{U^{\ast }V^{\ast }}$ is bounded between $\sigma
_{UV}$ and $\sigma _{UV}+\theta _{01}\overline{\sigma }_{\varepsilon }^{2}$.

Notice that, by construction, $\left[ \underline{\sigma }_{U^{\ast
}}^{2},\sigma _{U}^{2}\right]$ is a valid identified set, i.e., the
requirement $\sigma_{U^*}^2 \in \left[ \underline{\sigma }_{U^{\ast
}}^{2},\sigma _{U}^{2}\right]$ is necessary: any $\sigma_{U^*}^2 \notin %
\left[ \underline{\sigma }_{U^{\ast }}^{2},\sigma _{U}^{2}\right]$ would
violate at least one of the necessary primitive requirements considered
above.

Next we show that the constructed identified set for $\sigma _{U^{\ast
}}^{2} $ is sharp, i.e., for any $\sigma _{U^{\ast }}^{2}\in \lbrack 
\underline{\sigma }_{U^{\ast }}^{2},\sigma _{U}^{2}]$ there exist compatible 
$\sigma _{V^{\ast }}^{2}\geq 0$, $\sigma _{\varepsilon }^{2}\geq 0$ and $%
\sigma _{U^{\ast }V^{\ast }}$ satisfying $\sigma _{U^{\ast }V^{\ast
}}^{2}\leq \sigma _{U^{\ast }}^{2}\sigma _{V^{\ast }}^{2}$ consistent with
the distribution of the observables, i.e., such that equation~(\ref%
{eq:sigma2 w and wo stars}) holds. First, notice that if $\theta _{01}=0$, $%
\underline{\sigma }_{U^{\ast }}^{2}=\sigma _{U}^{2}$, so the identified set
for $\underline{\sigma }_{U^{\ast }}^{2}$is a singleton and, hence, it is
sharp. Suppose that $\theta _{01}\neq 0$. Consider any $\sigma _{U^{\ast
}}^{2}\in \lbrack \underline{\sigma }_{U^{\ast }}^{2},\sigma _{U}^{2}]$.
Solving for $\sigma _{\varepsilon }^{2}$, $\sigma _{V^{\ast }}^{2}$ and $%
\sigma _{U^{\ast }V^{\ast }}$ from equation~(\ref{eq:sigma2 w and wo stars})
gives 
\begin{equation}
\sigma _{\varepsilon }^{2}=(\sigma _{U}^{2}-\sigma _{U^{\ast }}^{2})/\theta
_{01}^{2},\quad \sigma _{V^{\ast }}^{2}=\sigma _{V}^{2}-(\sigma
_{U}^{2}-\sigma _{U^{\ast }}^{2})/\theta _{01}^{2},\quad \sigma _{U^{\ast
}V^{\ast }}=\sigma _{UV}+(\sigma _{U}^{2}-\sigma _{U^{\ast }}^{2})/\theta
_{01}.  \label{eq:pf:implied sigma stars}
\end{equation}%
Together, these define a (valid) data-generating process~(\ref{eq:dgp Y true}%
)-(\ref{eq:ME}) that is observationally equivalent to the data-generating
process in equations~(\ref{eq:dgp Y wo stars gen})-(\ref{eq:dgp UV wo stars
gen}). Thus, the identified set $[\underline{\sigma }_{U^{\ast }}^{2},\sigma
_{U}^{2}]$ is sharp. \hfill $\square $

\bigskip

\subsection{Derivation of equations~(\protect\ref{eq:true APE Tobit}) and (\protect\ref{eq:true APE Probit})}
\begin{align*}
APE_{j}^{\text{Tob}}\left( \sigma _{U^{\ast }}^{2}\right) & =\mathrm{E}\left[
\Phi \left( \frac{\theta _{0}^{\prime }H_{i}^{\ast }}{\sigma _{U^{\ast }}}%
\right) \theta _{0j}\right] =\mathrm{E}\left[ \Phi \left( \frac{\theta
_{01}X_{i}^{\ast }+\theta _{02}^{\prime }W_{i}}{\sigma _{U^{\ast }}}\right) %
\right] \theta _{0j}{} \\
& =\mathrm{E}\left[ \Phi \left( \frac{\theta _{01}\pi _{01}^{\prime
}Z_{i}+\left( \theta _{01}\pi _{02}+\theta _{02}\right) ^{\prime }W_{i}}{%
\sigma _{U^{\ast }}}+\frac{\theta _{01}V_{i}^{\ast }}{\sigma _{U^{\ast }}}%
\right) \right] \theta _{0j} \\
& =\mathrm{E}\left[ \Phi \left( \frac{\theta _{01}\pi _{01}^{\prime
}Z_{i}+\left( \theta _{01}\pi _{02}+\theta _{02}\right) ^{\prime }W_{i}}{%
\sigma _{U^{\ast }}\sqrt{1+\theta _{01}^{2}\sigma _{V^{\ast }}^{2}/\sigma
_{U^{\ast }}^{2}}}\right) \right] \theta _{0j} \\
& ={}\mathrm{E}\left[ \Phi \left( 
\frac{\theta _{01}\pi _{01}^{\prime }Z_{i}+\left( \theta _{01}\pi
_{02}+\theta _{02}\right) ^{\prime }W_{i}}{\sqrt{2\sigma _{U^{\ast
}}^{2}-\sigma _{U}^{2}+\theta _{01}^{2}\sigma _{V}^{2}}}\right) \right]
\theta _{0j},
\end{align*}%
where the penultimate equality comes from taking expectation with respect to 
$V_{i}^{\ast }$, and the last equality follows from equation~(\ref{eq:sigma2
w and wo stars}).

\bigskip

Derivation of equation~(\ref{eq:true APE Probit}) is similar:%
\begin{align*}
APE_{j}^{\text{Pr}}\left( \sigma _{U^{\ast }}^{2}\right) & =\mathrm{E}\left[
\phi \left( \frac{\theta _{0}^{\prime }H_{i}^{\ast }}{\sigma _{U^{\ast }}}%
\right) \frac{\theta _{0j}}{\sigma _{U^{\ast }}}\right] =\mathrm{E}\left[
\phi \left( \frac{\theta _{01}\pi _{01}^{\prime }Z_{i}+\left( \theta
_{01}\pi _{02}+\theta _{02}\right) ^{\prime }W_{i}+\theta _{01}V_{i}^{\ast }}{\sigma _{U^{\ast }}}%
\right) \right] \frac{%
\theta _{0j}}{\sigma _{U^{\ast }}} \\
& =\frac{1}{\sqrt{1+\theta _{01}^{2}\sigma _{V^{\ast }}^{2}/\sigma _{U^{\ast
}}^{2}}}\mathrm{E}\left[ \phi \left( \frac{\theta _{01}\pi _{01}^{\prime
}Z_{i}+\left( \theta _{01}\pi _{02}+\theta _{02}\right) ^{\prime }W_{i}}{%
\sigma _{U^{\ast }}\sqrt{1+\theta _{01}^{2}\sigma _{V^{\ast }}^{2}/\sigma
_{U^{\ast }}^{2}}}\right) \right] \frac{\theta _{0j}}{\sigma _{U^{\ast }}} \\
& ={}\mathrm{E}\left[ \phi
\left( \frac{\theta _{01}\pi _{01}^{\prime }Z_{i}+\left( \theta _{01}\pi
_{02}+\theta _{02}\right) ^{\prime }W_{i}}{\sqrt{2\sigma _{U^{\ast
}}^{2}-\sigma _{U}^{2}+\theta _{01}^{2}\sigma _{V}^{2}}}\right) \right] 
\frac{\theta _{0j}}{\sqrt{2\sigma _{U^{\ast }}^{2}-\sigma _{U}^{2}+\theta
_{01}^{2}\sigma _{V}^{2}}}.\qquad\quad \square
\end{align*}

\bigskip

\subsection{Proof of Proposition~\protect\ref{prop:IV-Mixed-Tobit ID}}

First, parameters $\pi _{0}$ are immediately identified from the first stage
regression.

Let $Q_{i}\equiv \pi _{01}^{\prime }Z_{i}+\pi _{02}^{\prime }W_{i}$, so $%
X_{i}=Q_{i}+V_{i}$. Since $\pi_0$ is identified, $Q_i$ is effectively
observed. First, notice that the p.d.f. of the joint distribution of $Y_i^*$
and $X_i$ given $Q_i$ and $W_i$ is given by 
\begin{eqnarray*}
f_{Y^{\ast }X|QW}\left( y,x|q,w\right) &=&f_{Y^*V|QW}\left( y,x-q|q,w\right)
\\
&=&f_{UV}\left( y-\left( \theta _{01}\left( q+v\right) +\theta _{02}^{\prime
}w\right) ,v|q,w\right) |_{v=x-q} \\
&=&f_{UV}\left( y-\left( \theta _{01}x+\theta _{02}^{\prime }w\right)
,x-q\right) \\
&=&\sum_{k=1}^{K}p_{k}\phi (y-\left( \theta _{01}x+\theta _{02}^{\prime
}w\right) ,x-q;\mu _{UV,k},\Sigma _{UV,k}).
\end{eqnarray*}%
Then,%
\begin{equation*}
f_{YX|QW}\left( y,x|q,w\right) =\left\{ 
\begin{array}{c}
f_{Y^{\ast }X|QW}\left( y,x|q,w\right) \text{ if }y>0; \\ 
\int_{-\infty }^{0}f_{Y^{\ast }X|QW}\left( t,x|q,w\right) dt\text{ if }y=0.%
\end{array}%
\right.
\end{equation*}

Consider the part of function $f_{YX|QW}\left( y,x|q,w\right) $ for $y>0$.
Since all the involved variables are observed, this function is identified
nonparametrically for all $y>0$ and $x,q,w$ (in the support of the
respective random variables). Since function $f_{Y^{\ast }X|QW}\left(
y,x|q,w\right) $ is entire (for any fixed $x,q,w$), identification of $%
f_{YX|QW}\left( y,x|q,w\right) $ for $y>0$ implies that $f_{Y^{\ast
}X|QW}\left( y,x|q,w\right) $ is identified for all $y$ and $x,q,w$.

Thus, identification of the model using the data on $\left(
Y_{i},X_{i},Z_{i},W_{i}\right) $ is equivalent to identification of the
model using the data on $\left( Y_{i}^{\ast },X_{i},Z_{i},W_{i}\right) $.
Then, $\theta _{0}$ is identified by the linear IV\ regression argument.
Hence, $U_i = Y_i^* - \theta_{01} X_i - \theta_{02}^{\prime }W_i$ and $V_i =
X_i - Q_i$ are effectively observed, and their joint density $f_{UV}\left(
u,v\right) $ (without any truncation) is identified, which in turn implies
that $\left\{ p_{k},\mu _{UV,k},\Sigma _{UV,k}\right\} _{k=1}^{K}$ are
identified.\hfill $\square $

\bigskip

\subsection{Proof of Proposition~\protect\ref{prop:general Ustar bounds}}

First, notice that $\mathcal{I}$ is a valid identified set, i.e., $%
\sigma_{U^*}^2 \in \mathcal{I}$ is a necessary requirement. Indeed, in the
proof of Proposition~\ref{prop:Ustar bounds}, we demonstrated that $%
\sigma_{U^*}^2 \in \mathcal{I}_{k}$ is a necessary requirement (notice that
the same argument applies because $\Sigma_{UV,k}$ is equal to $\Sigma_{UV,j
\ell}$ for some $j$ and $\ell$ satisfying the requirements in equation %
\eqref{eq:general reduced form to structural relationship}). Since this has
to hold for all $k \in \{1, \ldots, K\}$, $\sigma_{U^*}^2 \in \bigcap_{k} 
\mathcal{I}_{k} = \mathcal{I} $ is also a necessary requirement.

Next, we demonstrate that $\mathcal{I}$ is sharp. As in the proof of
Proposition~\ref{prop:Ustar bounds}, if $\theta _{01}=0$, the identified set
is a singleton, and the statement is trivial. Below, we consider $\theta
_{01}\neq 0$.

Let $\left\{ p_{V^*,j0},\mu _{U^{\ast } V^*,j0},\Sigma _{U^{\ast
}V^*,j0}\right\} _{j=1}^{J}$ and $\left\{ p_{\varepsilon, \ell 0},\mu
_{\varepsilon ,\ell 0},\sigma _{\varepsilon ,\ell 0}^{2}\right\} _{\ell
=1}^{L}$ denote the true values of the (structural) parameters that map into
the joint distribution of $(U_i,V_i)$ according to equation~%
\eqref{eq:general reduced
form to structural relationship}.

To demonstrate that $\mathcal{I}$ is sharp, we will show that for any $%
\sigma _{U^{\ast }}^{2}\in \mathcal{I}$, there exist a compatible set of
parameters $\left\{ p_{V^*,j},\mu _{U^{\ast } V^*,j},\Sigma _{U^{\ast
}V^*,j}\right\} _{j=1}^{J}$ and $\left\{ p_{\varepsilon, \ell},\mu
_{\varepsilon ,\ell},\sigma _{\varepsilon ,\ell}^{2}\right\} _{\ell =1}^{L}$
mapping into the same distribution of $(U_i,V_i)$ as the true parameters and
hence producing the same distributions of observables.

In, particular we will start with showing that for any $\sigma _{U^{\ast
}}^{2}\in \mathcal{I}$ there exist compatible $\sigma _{V^{\ast },j}^{2}\geq
0$, $\sigma _{\varepsilon ,\ell }^{2}\geq 0$, and $\sigma _{U^{\ast }V^{\ast
},j}$ satisfying $\sigma _{U^{\ast }V^{\ast },j}^{2}\leq \sigma _{U^{\ast
}}^{2}\sigma _{V^{\ast },j}^{2}$ mapping into the same collection of $%
\Sigma_{UV,j \ell}$ for $j\in \{1,\ldots ,J\}$ and $\ell \in \{1,\ldots ,L\}$%
.

Take any $\sigma _{U^{\ast }}^{2}\in \mathcal{I}$. First, notice that, for
any $\left( j,\ell \right) $, there exists $k(j,\ell)$ such that $%
\Sigma_{k(j,\ell)} = \Sigma_{UV,j \ell}$ satisfying the requirements in
equation \eqref{eq:general reduced form to structural relationship}. Since $%
\sigma_{U^*}^2 \in \mathcal{I}_{k(j,\ell)}$, there exist unique $\sigma
_{V^{\ast },j\ell }^{2}$, $\sigma _{\varepsilon ,j\ell }^{2}$, and $\sigma
_{U^{\ast }V^{\ast },j\ell }$ consistent with $\Sigma _{UV,j\ell } =
\Sigma_{k(j,\ell)}$ (as demonstrated in the proof of Proposition~\ref%
{prop:Ustar bounds}). This triplet is computed using equation~(\ref%
{eq:pf:implied sigma stars}) for any $\left( j,\ell \right) $.

Next, we need to show that these triplets are also internally consistent
along the $j$ and $\ell $ dimensions, i.e., that $\sigma _{V^{\ast },j\ell
}^{2}=\sigma _{V^{\ast },j\ell ^{\prime }}^{2}=\sigma _{V^{\ast },j}^{2}$
and $\sigma _{U^{\ast }V^{\ast },j\ell }=\sigma _{U^{\ast }V^{\ast },j\ell
^{\prime }}=\sigma _{U^{\ast }V^{\ast },j}$ for any $\ell ,\ell ^{\prime
}\in \{1,\ldots ,L\}$, and, similarly, $\sigma _{\varepsilon ,j\ell
}^{2}=\sigma _{\varepsilon ,j^{\prime }\ell }^{2}=\sigma _{\varepsilon ,\ell
}^{2}$ for any $j,j^{\prime }\in \{1,\ldots ,J\}$. Notice that according to
equation~(\ref{eq:general reduced form to structural relationship}), we also
have 
\begin{equation*}
\begin{pmatrix}
\sigma _{U,j\ell }^{2} & \sigma _{UV,j\ell } \\ 
\sigma _{UV,j\ell } & \sigma _{V,j\ell }^{2}%
\end{pmatrix}%
=%
\begin{pmatrix}
\sigma _{U^{\ast }0}^{2}+\theta _{01}^{2}\sigma _{\varepsilon ,\ell 0}^{2} & 
\sigma _{U^{\ast }V^{\ast },j0}-\theta _{01}\sigma _{\varepsilon ,\ell 0}^{2}
\\ 
\sigma _{U^{\ast }V^{\ast },j0}-\theta _{01}\sigma _{\varepsilon ,\ell 0}^{2}
& \sigma _{V^{\ast },j0}^{2}+\sigma _{\varepsilon ,\ell 0}^{2}%
\end{pmatrix}%
.
\end{equation*}%
Using the relationship above and the expressions for $\sigma _{V^{\ast
},j\ell }^{2}$ and $\sigma _{U^{\ast }V^{\ast },j\ell }$ obtained in the
proof of Proposition~\ref{prop:Ustar bounds}, for any $\ell $ we have 
\begin{gather*}
\sigma _{V^{\ast },j\ell }^{2}=\sigma _{V,j\ell }^{2}-(\sigma _{U,j\ell
}^{2}-\sigma _{U^{\ast }}^{2})/\theta _{01}^{2}=\sigma _{V^{\ast
},j0}^{2}-(\sigma _{U^{\ast }0}^{2}-\sigma _{U^{\ast }}^{2})/\theta
_{01}^{2}=\sigma _{V^{\ast },j}^{2}, \\
\sigma _{U^{\ast }V^{\ast },j\ell }=\sigma _{UV,j\ell }+(\sigma _{U,j\ell
}^{2}-\sigma _{U^{\ast }}^{2})/\theta _{01}=\sigma _{U^{\ast }V^{\ast
},j0}+(\sigma _{U^{\ast }0}^{2}-\sigma _{U^{\ast }}^{2})/\theta _{01}=\sigma
_{U^{\ast }V^{\ast },j},
\end{gather*}%
where the last equalities in both lines provide consistent definitions of $%
\sigma _{V^{\ast },j}^{2}$ and $\sigma _{U^{\ast }V^{\ast },j}$, which do
not depend on $\ell $. 
Similarly, for any $j$, we have 
\begin{equation*}
\sigma _{\varepsilon ,j\ell }^{2}=(\sigma _{U,j\ell }^{2}-\sigma _{U^{\ast
}}^{2})/\theta _{01}^{2}=\sigma _{\varepsilon ,\ell 0}^{2}+(\sigma _{U^{\ast
}0}^{2}-\sigma _{U^{\ast }}^{2})/\theta _{01}^{2}=\sigma _{\varepsilon ,\ell
}^{2},
\end{equation*}%
where the last equality defines $\sigma _{\varepsilon ,\ell }^{2}$, which
does not depend on $j$.

Finally, notice that, by construction, $\left\{ p_{V^{\ast },j0},\mu
_{U^{\ast }V^{\ast },j0},\Sigma _{U^{\ast }V^{\ast },j}\right\} _{j=1}^{J}$
and $\left\{ p_{\varepsilon ,\ell 0},\mu _{\varepsilon ,\ell 0},\sigma
_{\varepsilon ,\ell }^{2}\right\} _{\ell =1}^{L}$ map into the same joint
distribution of $(U_{i},V_{i})$ as the true parameters $\left\{ p_{V^{\ast
},j0},\mu _{U^{\ast }V^{\ast },j0},\Sigma _{U^{\ast }V^{\ast },j0}\right\}
_{j=1}^{J}$ and $\left\{ p_{\varepsilon ,\ell 0},\mu _{\varepsilon ,\ell
0},\sigma _{\varepsilon ,\ell 0}^{2}\right\} _{\ell =1}^{L}$. Since we
picked an arbitrary $\sigma _{U^{\ast }}^{2}\in \mathcal{I}$, this completes
the proof that $\mathcal{I}$ is sharp.\hfill $\square $

\end{appendices}%

\end{document}